\documentclass[]{aa}  

\usepackage{graphicx}
\usepackage{adjustbox}
\usepackage{txfonts}
\usepackage{natbib}

\newcommand{\Teff}{$T_\mathrm{eff}$}

\usepackage{subcaption}
\usepackage{caption}     

\begin{document}

   \title{A Large-Sample Study of Interstellar Reddening in Galactic Star Clusters}

  \author{Ernst Paunzen\inst{1}
   \and Teja Begari\inst{2,3}}

\institute{Department of Theoretical Physics and Astrophysics, Faculty of Science, Masaryk University, Kotl\'{a}\v{r}sk\'{a} 2, 611 37 Brno, Czechia
     \email{epaunzen@physics.muni.cz}
     \and Department of Physics, Dr. Vishwanath Karad MIT World Peace University, Pune, India
     \and American Association of Variable Star Observers, 185 Alewife Brook Parkway, Suite 410, Cambridge, MA 02138, USA}

  \abstract
   {The interstellar reddening (extinction or absorption) is caused mainly by the dust grains of different
   sizes and compounds. All astrophysical observations are somehow affected by this effect, depending
   on the wavelength and location of the observed object. Different methods and techniques have been developed
   over the last nearly one hundred years to determine the reddening and to map the solar neighbourhood.}
   {The members of open clusters are a consistent population of stars formed together with a homogeneous
   metallicity. Therefore, the reddening can be estimated in a statistical way, avoiding all problems
   of single stars and a spectral-energy-distribution fitting, for example. Other than for most available
   works, we can also use more massive upper main-sequence stars for the analysis.}
   {We used classical colour-colour diagrams (also known as Q-method) for $(B-V)$, $(BP-RP)$, $(U-B)$, 
   $(U-BP)$, $(U-RP)$, and $(U-G)$ to determine consistent mean reddening values and their errors. The
   corresponding standard lines for the fitting are independent of the age and metallicity.}
   {We determined colour excesses for 5377 Galactic open clusters covering Galactocentric distances 
   from two to twenty kpc, respectively. The
   derived values were checked with existing dust and extinction maps. In general, we found a
   good agreement. The solar neighbourhood of $\pm$1500\,pc shows a very patchy structure of the 
   dust composition.}
   {Our results lay the foundations for constructing a new map of the reddening (dust) distribution 
   and estimating clusters' metallicities by isochrone fitting keeping the distances and 
   reddening values fixed.}

   \keywords{open clusters and associations: general -- dust, extinction -- Techniques: photometric -- Hertzsprung-Russell and C-M diagrams}

   \maketitle
%

\section{Introduction} \label{introduction} 

Interstellar reddening, caused by wavelength-dependent extinction by dust grains in the interstellar medium (ISM), plays a fundamental role in shaping the observed properties of Galactic stellar populations
\citep{2003ARA&A..41..241D}. Accurate reddening determination is essential for deriving reliable stellar and cluster parameters, including distances, ages, and metallicities. Open clusters, owing to their well-defined stellar populations and wide spatial distribution across the Galactic disk, provide powerful tracers of the large-scale structure of interstellar dust and extinction \citep{2020MNRAS.499.1874M}.

We want to emphasise that the expressions reddening, absorption, and reddening are all used for the same
theoretical and observational phenomenon: the loss of photons radiated by an astrophysical object and the
observer \citep{2004AJ....128.2144M}.

Traditionally, reddening in open clusters has been estimated using colour–magnitude diagrams (CMDs) and two-colour diagrams, where the displacement of cluster sequences from intrinsic loci allows the colour excess to be inferred. By fitting isochrones, not only the reddening, but also the distance, metallicity and age are
additional free parameters. This makes the overall automatic fitting process error-prone, especially for
non-standard cases \citep{2015A&A...582A..19N}. Even new and accurate data from the $Gaia$ satellite have
not improved the situation \citep[for example, see Figure 6 in][]{2025AJ....170..288L}.

Among classical approaches, the Q-method \citep{1975PASP...87..425G} has been widely employed to derive reddening independently of distance by combining multiple photometric colours, particularly for early-type stars. This method is most powerful using a filter in the Ultraviolet \citep{1958LowOB...4...37J}.
With the advent of large photometric surveys and astrometric data, CMD-based techniques have been significantly refined, enabling homogeneous reddening estimates for large samples of clusters.

In addition, reddening ratios like $E(U-B)/E(B-V)$ trace the composition and characteristics of the 
ISM on small and large scales \citep{2001ApJ...548..296W}. Such knowledge is important for analysing,
for example, the star-forming history of our Milky Way \citep{2020ARA&A..58..157T}. Such maps exist
based on the Galactic field stars. One of the most severe shortcomings of these efforts is
the usage of only lower main sequence objects, i.e. stars with \Teff\,$<$\,7000\,K \citep{2024ApJS..272...20A}.
Such works completely miss the young and high-mass population of our Milky Way, which introduces 
a bias towards old stars. 

In recent years, several extensive compilations of Galactic open clusters have become available, providing unprecedented opportunities for statistical studies of cluster properties and potential members
\citep{2023MNRAS.526.4107P, 2024A&A...686A..42H}. In particular, the catalogue by \cite{2024A&A...686A..42H} presents homogeneous determinations of fundamental parameters for a large number of open clusters using $Gaia$ photometry and astrometry, including estimates of colour excess and extinction. Complementary efforts based on $Gaia$ data and multi-band photometry have further expanded cluster catalogues and improved parameter precision, allowing systematic investigations of Galactic structure and interstellar extinction on kiloparsec scales.

As stated before, the applied Q-method is most efficient using a filter in the Ultraviolet, such as
Johnson $U$. Such observations are hardly performed any more, as only a very few publications
in recent years have presented
corresponding data for open clusters \citep{2023AJ....165...79Y}. The reasons are manifold: the insensitivity
of most currently used detectors and the influence of the sky background of the observation site, just two examples.
Therefore, for this work, we chose to present a pioneering study using synthetic
Johnson $UBV$ (based on $Gaia$ DR3, BP/RP spectra) 
and $Gaia$ $GBPRP$ photometry for a large-sample study of interstellar reddening in 
Galactic open clusters.

\begin{figure}
    \centering
    \includegraphics[width =0.95 \columnwidth]{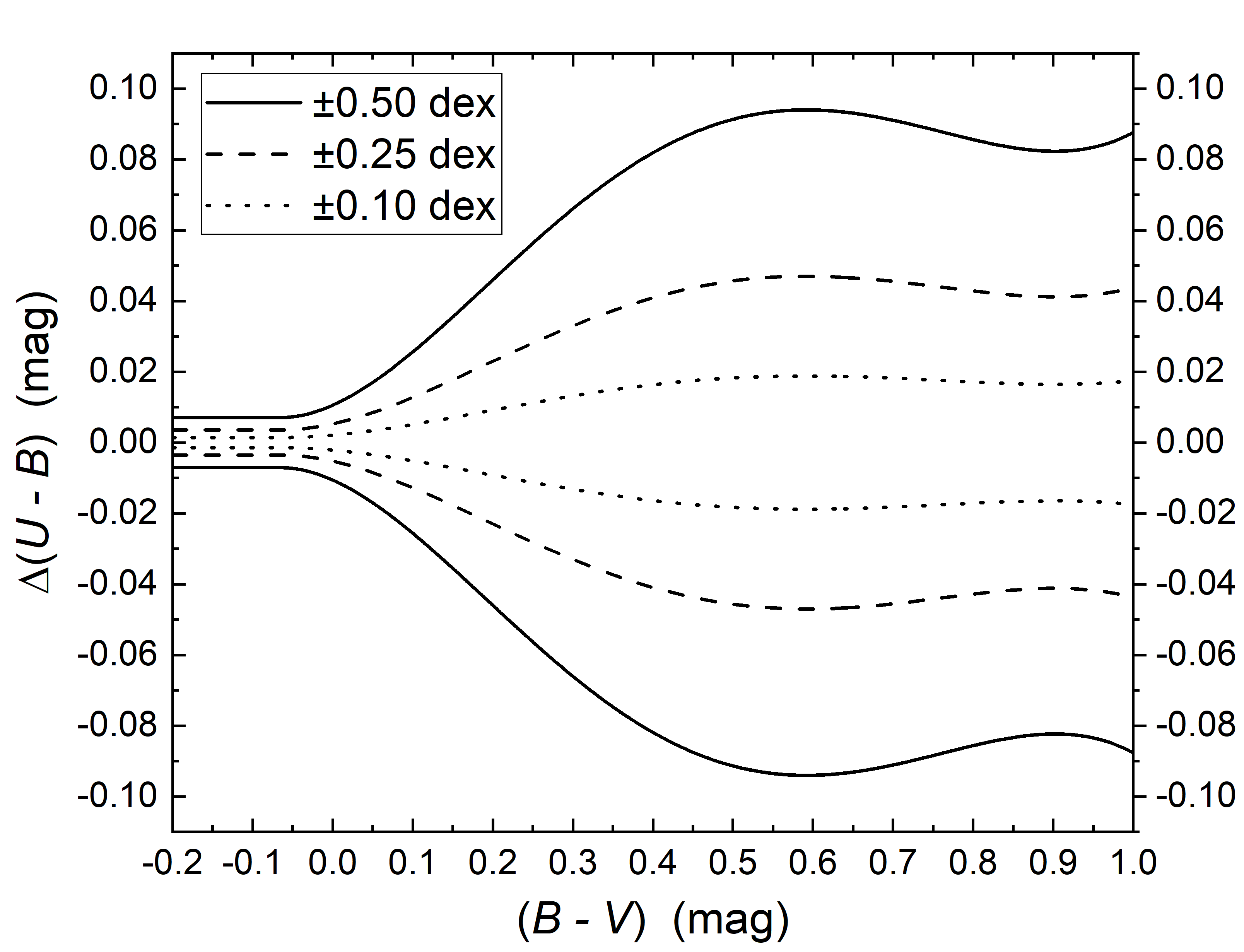}
    \caption{The effect of metallicity on the standard lines (essentially the 
    Zero Age Main Sequences) of $\Delta(U-B)$ versus $(B-V)$ for the range found for the clusters 
    studied here. A value of $\Delta(U-B)$\,=\,0 corresponds to the solar metallicity. 
    \citet{2022MNRAS.509..421N} showed that 98\% of all open clusters within 2000\,pc around the
    Sun have metallicities of $\pm$0.25\,dex, respectively.}
    \label{metallicity_BV_UB}
\end{figure}

\begin{figure}
    \centering
    \includegraphics[width =0.95 \columnwidth]{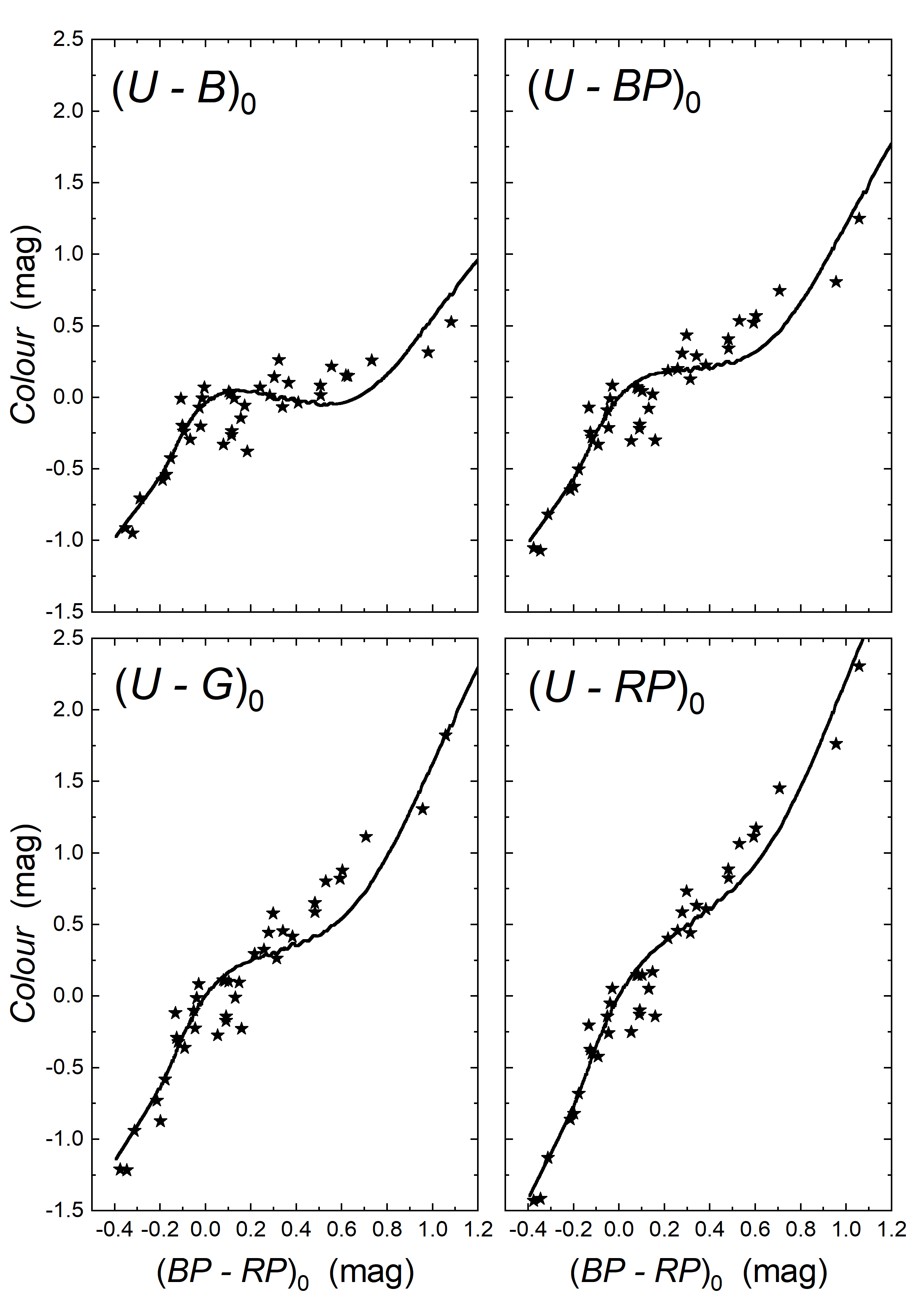}
    \caption{The colour-colour diagrams for the young open cluster NGC 2244, which
    shows severe differential reddening ($\Delta A_\mathrm{V}$\,=\,0.95\,mag). The scales for
    all axes are identical. The shapes of the standard lines over the whole spectral range 
    make the applied diagnostic tool almost independent of the effects of differential reddening.}
    \label{NGC2244}
\end{figure}

\begin{figure}
    \centering
    \includegraphics[width =0.95 \columnwidth]{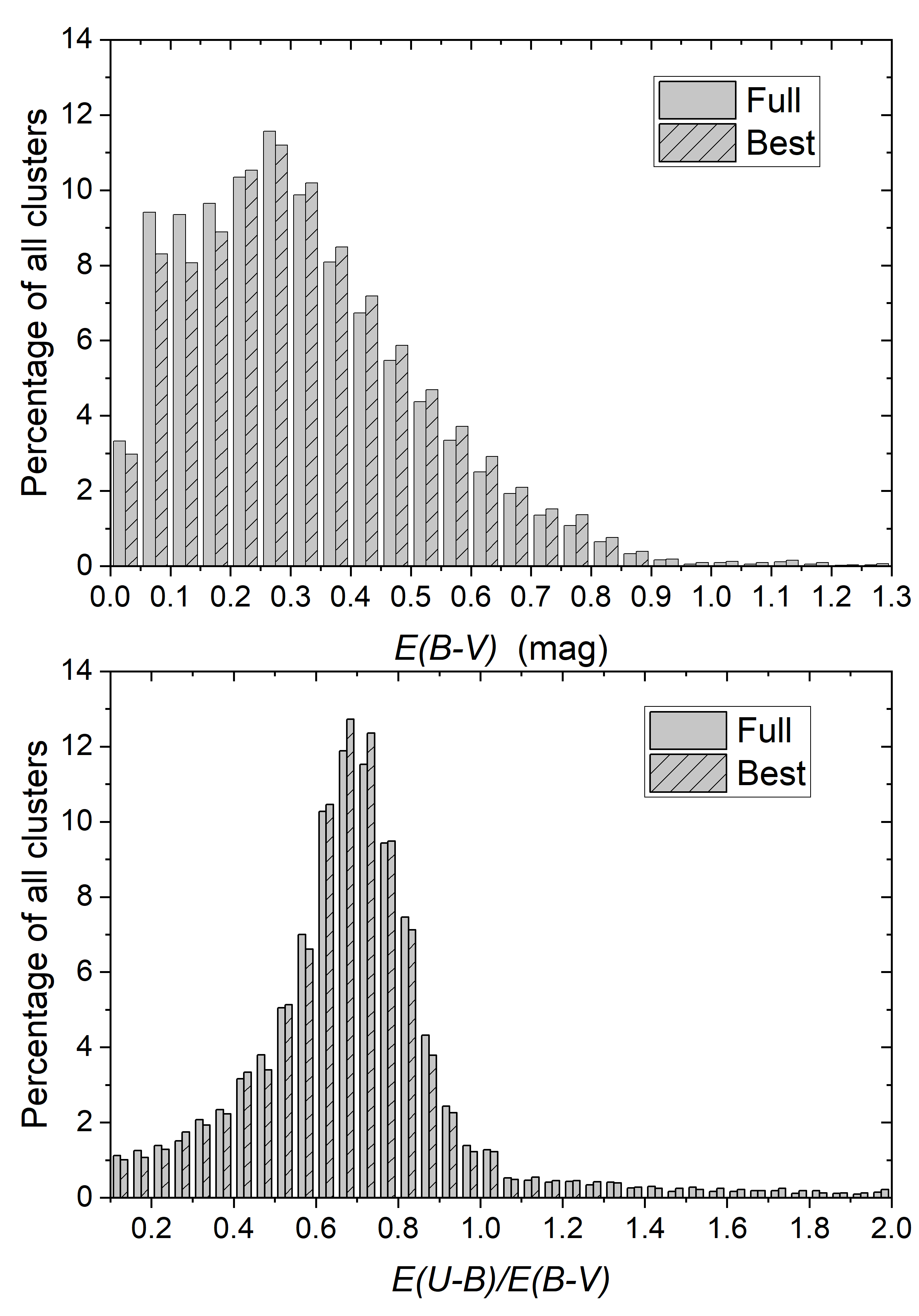}
    \caption{Distributions of the colour excess $E(B-V)$ (upper panel) and the reddening ratio $E(U-B)/E(B-V)$
    (lower panel). We omitted 
    very large $E(U-B)/E(B-V)$ values due to small $E(U-B)$ values. We show the full sample and only 
    ``the best'' defined open clusters according to the criteria by \citet{2023MNRAS.526.4107P}. The distributions show no statistical differences.}
    \label{ReddeningVsNo-clusters}
\end{figure}

\section{Target selection and synthetic Johnson $UBV$ photometry}

First of all, we excluded all known Globular clusters from the lists by 
\citet{2010arXiv1012.3224H} and \citet{2021MNRAS.505.5957B}.
We used the membership list of open clusters from the following two references.

\textit{\citet{2023MNRAS.526.4107P}\footnote{\url{https://ucc.ar/}}}: They used FASTMP (acronym for Fast Membership Probabilities)
for estimating membership probabilities. It can run in supervised and unsupervised modes.
It is based on pyUPMASK \citep{2021A&A...650A.109P}. 

\textit{\citet{2023A&A...673A.114H}}: They applied the Hierarchical Density-Based Spatial Clustering 
of Applications with Noise (HDBSCAN) algorithm to recover star clusters. 
They validated the found aggregates using a statistical density test and a Bayesian convolutional 
neural network for colour-magnitude diagram classification. 

The identification of star cluster members, and even of star clusters themselves, is not straightforward.
We just want to recall that the basic definition of a star cluster is that the members are born in the same
molecular cloud and are held together by gravity. With the discovery of thousands of new Moving Groups 
\citep{2018ApJ...863...91F}, the above-mentioned definition becomes utterly important. The boundaries 
are blurring. The two used sources attempted to avoid, or at least flag, Moving Groups. Another problem
is the unambiguous identification of star clusters. Several papers claimed the detection of new star clusters,
which turned out to be already known or just a detection of a part of a much larger aggregate. We matched both
sources, deleted duplicates, and skipped open clusters that could not be unambiguously verified.
We identified the open clusters using not only the name, but most importantly, the central coordinates
and given mean proper motions.
Then we merged both lists using members which are given with a probability of more than fifty per cent 
in at least one reference.

The $Gaia$ DR3 provides flux-calibrated low-resolution spectrophotometry (BP/RP spectra) 
for about 200 million sources in the wavelength range from
330 to 1050\,nm \citep{2023A&A...674A..33G}.
From these spectra, synthetic photometry can be derived for any passband. It is claimed that
existing observations can be reproduced within a few per cent over a wide range of 
magnitudes and colour for wide and medium bands, and with up to millimag accuracy 
when synthetic photometry is standardised for external sources.
In Section 3.2. of \citet{2023A&A...674A..33G} the standardisation of the 
Johnson-Kron-Cousins system is described. Another issue is the standardisation of Johnson $U$
because there is a colour-dependent systematic error affecting the UV region bluer than 400\,nm
\citep{2023A&A...674A...3M}. We have therefore checked the synthetic Johnson $UBV$ photometry using
the mean observed data from the General Catalogue of 
Photometric Data \citep[GCPD\footnote{\url{https://gcpd.physics.muni.cz/}}][]{1997A&AS..124..349M}.
In total, we used 30\,377 measurements for stars with 3.0\,$<$\,$V$\,$<$\,17.7\,mag, respectively.
We then calculated $\Delta U$\,=\,$U_\mathrm{synth}$\,$-$\,$U_\mathrm{obs}$, $\Delta B$, and
$\Delta V$. The mean values and the standard deviations
are $-$0.03(8)\,mag, $-$0.01(6)\,mag, and +0.02(6)\,mag, respectively. We therefore conclude that
the synthetic Johnson $UBV$ photometry by the $Gaia$ consortium can be used without any
further corrections.

\begin{figure}
    \centering
    \includegraphics[width=0.95\linewidth]{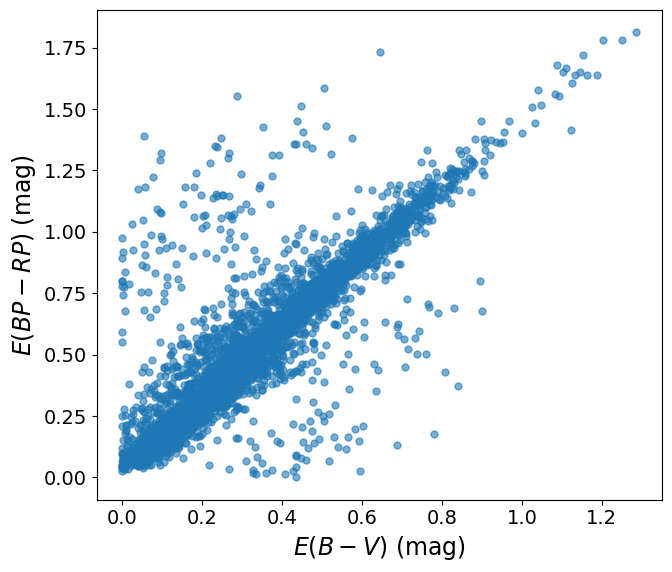}
    \caption{Comparison between $E(B-V)$ and $E(BP-RP)$ for the full cluster sample. As expected, a 
    strong positive correlation is evident between the two colour excess estimates. There are also 
    several outliers visible.}
    \label{fig:comparison}
\end{figure}

\begin{figure}
    \centering
    \includegraphics[width=0.95\linewidth]{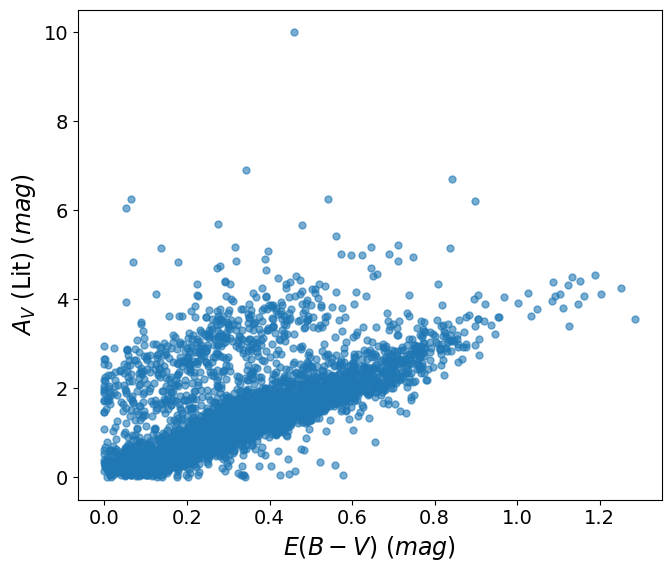}
    \caption{Comparison between our derived colour excess 
     $E(B-V)$ and the literature $A_\mathrm{V}$(Lit) values for the full cluster sample showing strong correlations with increasing scatter at higher reddening.}
    \label{fig:AVvsE}
\end{figure}

\section{Analysis} \label{analysis} 

It is well known that a classical colour-colour diagram $(U-B)$ versus $(B-V)$ is independent of the
luminosity and metallicity \citep{1978A&A....62..411B}. As we extended the analysis to the $Gaia$
DR3 photometric system, let us recall the filters 
and their characteristics\footnote{\url{http://svo2.cab.inta-csic.es/theory/fps/}}: $G$
($\lambda_\mathrm{mean}$\,=\,672\,nm, $W_\mathrm{eff}$\,=\,405\,nm), 
$BP$ (504, 216), and $RP$ (762, 292). They are much broader than the Johnson $UBV$ system
($W_\mathrm{eff}$ between 66 and 97\,nm). We used $(B-V)$ and $(BP-RP)$ versus $(U-B)$, 
$(U-BP)$, $(U-RP)$, and $(U-G)$.

For the corresponding standard lines (essentially the Zero Age Main Sequences), we used the 
non-rotating PARSEC V2.0 isochrones \citep{2025A&A...701A.258N} for [M/H]\,=\,0, respectively. 
Figure \ref{metallicity_BV_UB} shows the effect of metallicity on the standard 
lines of $\Delta(U-B)$ versus $(B-V)$ for the range found for the clusters 
studied here. A value of $\Delta(U-B)$\,=\,0 corresponds to the solar metallicity. 
\citet{2022MNRAS.509..421N} showed that 98\% of all open clusters within 2000\,pc around the
Sun have metallicities of $\pm$0.25\,dex, respectively. The absolute deviations 
($\pm$0.04\,mag for $\pm$0.25\,dex and spectral types cooler than F8) for the 
other colour-colour diagrams are similar. As expected,
the effect of metallicity of $\pm$0.5\,dex \citep[the range of open clusters' 
metallicities up to a galactocentric distance of 21\,kpc, ][]{2022MNRAS.509..421N} is negligible
taking into account the natural width of the main-sequence due to rotation, for example.

As the next step, we fitted the standard lines to all eight combinations 
of the colour-colour
diagrams for open clusters with at least seven members available. This limit was chosen to avoid
spurious results due to poor number statistics.
We shifted the x- and y-axes independently, without using any predefined vector. 
We perform the fitting process by minimising the distances of all members simultaneously.
As an alternative approach, we used weights proportional to the membership probabilities. 
Here we face two different problems. The sources used define the membership probabilities 
differently. So simply taking the mean values for stars included in both sources could
introduce an unknown bias. The second problem is that \citet{2023A&A...673A.114H} 
present a wide range of absolute membership probabilities. For example, most of the 
well-studied NGC clusters range from 0 to 1, respectively. For sparser open
clusters, for example, Alessi 53 (just to pick one), they range from 0.965 to 1, respectively.
This makes their direct use for different open clusters difficult. We calculated the
reddening values using the averaged mean membership probabilities as weights. For 90\%
of the sample, the results were consistent with the unweighted ones within 2\%, respectively. 
For the remaining 10\%, we saw some strange effects, probably due to the ill-defined
absolute range. Therefore, we used the unweighted reddening values in our analysis. 
Finally, we derived mean values and their standard deviations for the extinction in the different
colours. We have four diagrams including $(B-V)$ and $(BP-RP)$ and two diagrams with 
$(U-B)$, $(U-BP)$, $(U-RP)$, and $(U-G)$ (see Fig. \ref{NGC1912} for NGC 1912).
In total, we derived colour excesses for 5377 individual star clusters. 
They cover Galactocentric distances of 2 to 20 kpc. Examples of the fitted 
standard lines and the derived colour excesses can be seen in Figure \ref{BV_UB_12_OCLs}.
It shows the whole variety of number of members, ages, distances and reddening values.

One critical issue is the differential reddening of young open clusters \citep{2001MNRAS.328..370Y}.
This is caused by dust and gas from the initial molecular cloud from which the open cluster is formed.
In general, it affects open clusters and 
star-forming regions younger than 10\,Myr \citep{2001ApJ...562..852H}. The clusters with a differential
reddening $\Delta E(B-V)$ 
of almost 1\,mag taken from \citet{2001MNRAS.328..370Y} are NGC 2244, Trumpler 16, and Trumpler 37.
However, this large effect is not reflected by the mean visual absorptions $\Delta A_\mathrm{V}$ 
and their errors in the literature, which are 0.18, 0.07, and 0.09\,mag. We show the effect of differential
reddening using the data for NGC 2244 ($\Delta A_\mathrm{V}$\,=\,0.95\,mag).

Figure \ref{NGC2244} shows the colour-colour diagrams for $(BP-RP)_\mathrm{0}$ of 
NGC 2244. The shapes of the 
standard lines over the whole spectral range make the applied diagnostic tool almost independent 
of the effects of differential reddening.

Figure \ref{ReddeningVsNo-clusters} presents the histograms of 
the colour excess $E(B-V)$ (upper panel) and the reddening ratio $E(U-B)/E(B-V)$ 
(lower panel). 
When calculating extinction ratios, we must be careful when the divisor is very small.
Then we could end up with very large and unphysical values. Therefore, we omitted 
very large $E(B-V)/E(U-B)$ values due to small $E(U-B)$ values. The peak at zero values is caused
by values close to zero of both $E(B-V)$ and $E(U-B)$, respectively.
Let us recall the following relation for the standard reddening law  
($R_\mathrm{V} = 3.1$) was found \citep{1999PASP..111...63F}
\begin{equation}
E(U-B)/E(B-V) = 0.69 + 0.04\,E(B-V).    
\end{equation}
We find the peak of the distribution (Fig. \ref{ReddeningVsNo-clusters}) at 0.72, which is
the expected mean value for a range of $E(B-V)$ values in the solar neighbourhood.

We also plotted in Fig. \ref{ReddeningVsNo-clusters} the full sample and only 
``the best'' defined open clusters according to the criteria by \citet{2023MNRAS.526.4107P}. 
They comprise the C3 parameter, which is the combined C1 and C2 classes, described in Sect. 4.3 of 
the listed reference. Basically, they measure the contrast between the spatial distribution
of cluster members and that of the surrounding field stars within the 
five-dimensional parameter space defined by celestial coordinates, proper motions, and parallax.
Furthermore, the likelihood that the observed stellar sequence of the candidate members is
statistically indistinguishable from a sequence randomly drawn from the field star population is calculated.
This sample comprises 3286 open clusters. The distributions show no statistical differences. So, we 
continued using the full sample in the further analysis.

Table \ref{table_master} presents the essential data and results for our sample of star clusters. 
The full table is only available online. It is organised as follows:

\begin{itemize}
\item Column 1: Name.
\item Column 2: Distance from the Sun taken from the literature.
\item Column 3: Absorption in the $V$ band taken from the literature.
\item Column 4: Galactic longitude (J2000).
\item Column 5: Galactic latitude (J2000).
\item Column 6: X, in the direction of the Galactic centre.
\item Column 7: Y, in the direction of the Galactic disk rotation.
\item Column 8: Z, in the direction of the North Galactic Pole.
\item Column 9: Galactocentric distance.
\item Column 10: $E(B-V)$.
\item Column 11: $\sigma E(B-V)$.
\item Column 12: $E(BP-RP)$.
\item Column 13: $\sigma E(BP-RP)$.
\item Column 14: $E(U-B)$.
\item Column 15: $\sigma E(U-B)$.
\item Column 16: $E(U-BP)$.
\item Column 17: $\sigma E(U-BP)$.
\item Column 18: $E(U-G)$.
\item Column 19: $\sigma E(U-G)$.
\item Column 20: $E(U-RP)$.
\item Column 21: $\sigma E(U-RP)$.
\item Column 22: Number of used members.
\end{itemize}

\begin{figure*}
    \centering
    \begin{subfigure}{0.49\textwidth}
        \centering
        \includegraphics[width=\linewidth]{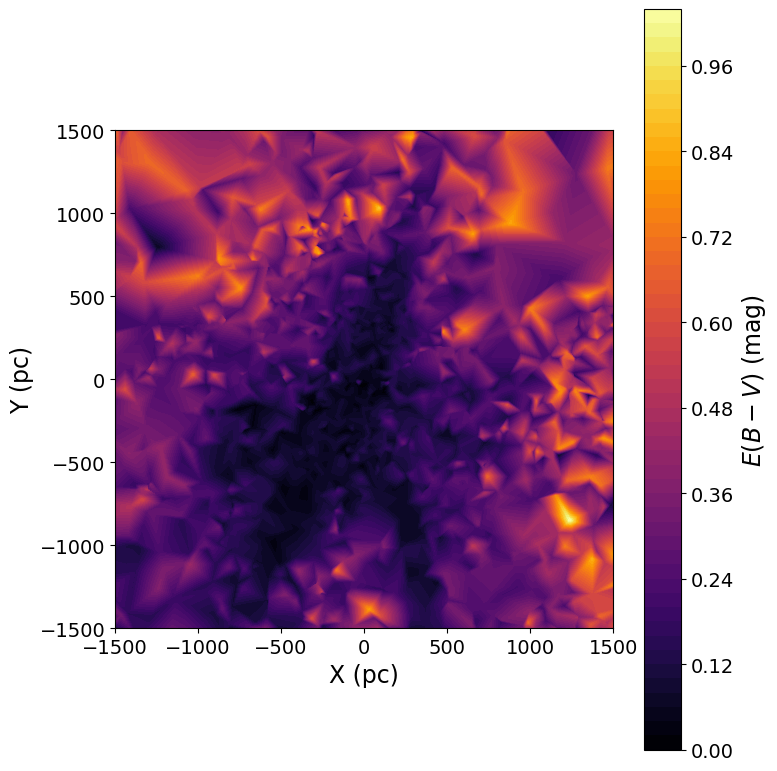}
    \end{subfigure}
    \hfill
    \begin{subfigure}{0.49\textwidth}
        \centering
        \includegraphics[width=\linewidth]{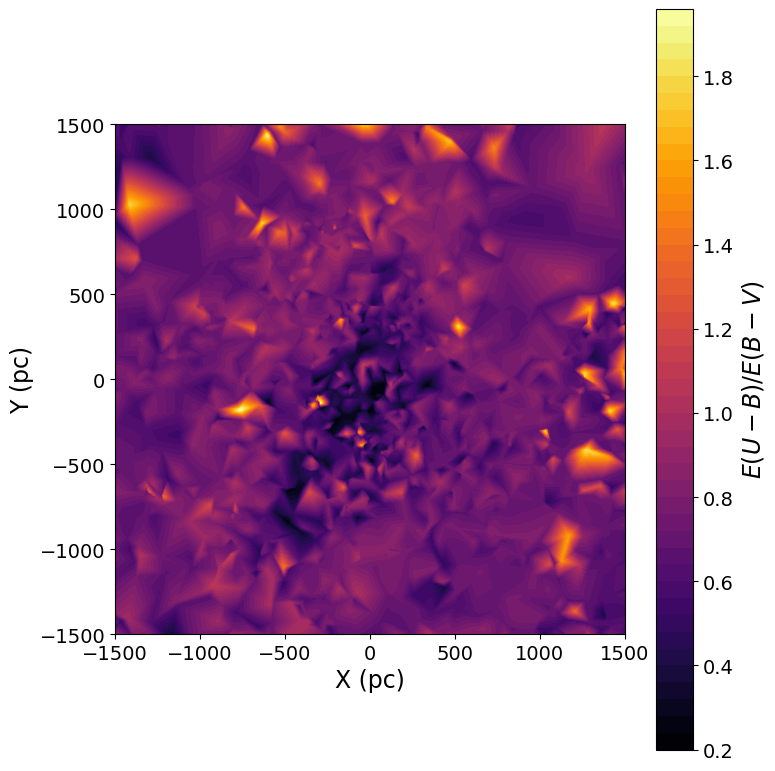}
    \end{subfigure}

    \caption{Spatial distribution of 2051 Galactic open clusters in the Galactic plane ($\pm125\,pc$)
    within $\pm$1500\,pc of the Sun, colour-coded by the reddening (left panel) and extinction
    ratio (right panel). Both maps reveal increasing reddening with distance from the Solar neighbourhood and localized high-extinction regions associated with interstellar dust structures. The Sun is at [0, 0].}
    \label{fig:XvsY}
\end{figure*}

Some of our values differ significantly from those reported in the literature
(Fig. \ref{fig:AVvsE}). Therefore, we
used our values and the mean distances calculated from the members' parallaxes of the
$Gaia$ DR3 to get CMDs. For this, we used synthetic Johnson $BV$ and $Gaia$ photometric 
magnitudes. For all deviating cases, we could fit the corresponding CMDs using our reddening values. In Fig. \ref{BV_MV_12_OCLs} we show a few examples.. We are therefore confident that our estimates are correct.

The transformation of extinction of the $Gaia$ passbands to $A_\mathrm{0}$ 
(the extinction at 550\,nm)
is not straightforward. The $Gaia$ consortium lists for the 
DR3\footnote{\tiny{\url{https://www.cosmos.esa.int/web/gaia/edr3-extinction-law}}}
a function of the star's intrinsic colour or effective temperature, with ten terms
for the 
extinction coefficient in the $m$ band ($k_\mathrm{m} = A_\mathrm{m}/A_\mathrm{0}$), respectively.
However, this formula is only valid for \Teff\,$<$\,10000\,K or stars cooler than A-type. 
To get a mean transformation coefficient for a star cluster consisting of hundreds of members
over the entire spectral range is only an approximation. Using this formula, we get
$k_\mathrm{G} = 0.84$, $k_\mathrm{BP} = 1.04$, and $k_\mathrm{RP} = 0.65$, respectively.
More often, the ratio with the absorption in Johnson $V$, $A_\mathrm{V}$ is 
listed \citep{2023ApJ...946...43W}. However, $A_\mathrm{V}$ and $A_\mathrm{0}$ are
very similar without significant offsets and correlations over the whole main sequence range
\citep{2022A&A...664A.174V}. If we take the standard reddening law by 
\citet{1999PASP..111...63F}\footnote{\tiny{\url{https://extinction.readthedocs.io/en/latest/}}}
for $R_\mathrm{V} = A_\mathrm{V}/E(B-V) = 3.1$, we get $\lambda_\mathrm{eff}$ for
$G$, $BP$, and $RP$ of 614, 527, and 726\,nm, respectively. These are close to the nominal listed
values.

In Fig. \ref{fig:comparison}, we present the comparison between $E(B-V)$ and $E(BP-RP)$ for 
the full cluster sample. As expected, a strong positive correlation is evident between the two 
colour excess estimates. But there are also several outliers evident. The colour-colour diagrams
of these open clusters were checked, and the derived colour excesses fit them very well. 
We also found no correlations with the Galactic location and age.
Reasons
for the non-standard coefficient could be an extreme reddening law or an extreme non-solar
metallicity.

We find the following correlation for our cluster sample ($N$\,=\,5377)
\begin{eqnarray}
    E(BP-RP) &=& 1.415(7)\,E(B-V) \\
    E(U-BP) &=& 1.878(15)\,E(U-B) \\
    E(U-G) &=& 2.667(22)\,E(U-B) \\
    E(U-RP) &=& 3.516(31)\,E(U-B) 
\end{eqnarray}
which includes aggregates at all different distances from the Sun.

As the next step, we investigated the solar 
neighbourhood ($\pm$1500\,pc) using Galactic [XYZ] coordinates. Let us recall that
X is in the direction of the Galactic centre, Y in the direction of the Galactic disk rotation, and Z
in the direction of the North Galactic Pole.
We used all clusters with Z\,=\,$\pm$125\,pc to represent the Galactic disk \citep{2025A&A...694A..70M}.
In total, this leaves us with 2051 star clusters from our sample. 

Figure \ref{fig:XvsY} shows the spatial distribution of the reddening (left panel) and extinction
ratio (right panel). We can immediately compare the distribution of $E(B-V)$ with the dust map
published by \citet[][Fig. 13 therein]{2022A&A...664A.174V}. There is excellent agreement of the 
almost dust-free regions in the second and third quadrants, as well as the high density line-of-sights.
Beyond a radius of about 500\,pc, the reddening becomes significantly higher. Looking at the 
extinction ratio, we see a very patchy structure which reflects the local structure of dust and gas
in the solar neighbourhood. Also, the Local Bubble \citep{2024ApJ...973..136O} is nicely 
reproduced. The features within both figures do not necessarily overlap because, on the left, we see 
the overall density, whereas on the right, we see the characteristics (composition) of the ISM.
This is the advantage of studying extinction ratios, including a filter in the
ultraviolet.

\section{Conclusions} \label{conclusions}

The study of the reddening and the extinction ratio, including a filter in the ultraviolet, is essential
to study the distribution and characteristics of the ISM. Available maps are mainly based on 
cool-type field stars, which introduces a bias by not including hot and rather young objects. Another
disadvantage is the use of spectral-energy-distribution fitting, which depends on an accurate
absolute flux calibration and the knowledge of the object's metallicity.

We present the results of fitting standard lines (independent of the metallicity and age)
to synthetic Johnson $UBV$ and $Gaia$ $GBPRP$ photometry of Galactic star clusters.
These are groups of stars born together with a common metallicity. Using 
$(B-V)$ and $(BP-RP)$ versus $(U-B)$, 
$(U-BP)$, $(U-RP)$, and $(U-G)$, we derived mean values and their errors. 
The sample of 5377 star clusters
covers Galactocentric distances from two to twenty kpc.

Let us recall that the four free parameters when fitting isochrones to colour-magnitude diagrams
of star clusters are the distance, metallicity, reddening, and age. Thanks to the $Gaia$ satellite,
we nowadays have a good knowledge of the mean distances based on parallax measurements. The age 
can be estimated by knowing the location of the main-sequence turn-off point and the absolute magnitude
of the cluster's red giants (if available). With precise knowledge of the reddening, the metallicity
can be estimated. Up to now, the published reddening values have deviated significantly for many star 
clusters, preventing to keep the metallicity as a free parameter. In such cases, it was kept fixed as
being solar.

We compared our results with existing reddening maps and found an excellent agreement. We could
trace the Local Bubble and other well-known features with the help of star clusters. The next steps are
now to construct a new map of the reddening distribution using our results and the ones already published.
However, this task is not straightforward because the already existing maps are not very homogeneous, with
significant offsets and inconsistencies. Finally, it is necessary to estimate
cluster parameters, especially the metallicities, by fitting isochrones, keeping the distances and 
reddening values fixed. This will significantly improve our knowledge about Galactic star clusters.


\begin{acknowledgements}

This research was carried out within the institutional support framework 
for the development of the Research Organisation of Masaryk University. 
This paper is dedicated to G. Gebhard, who died during its preparation.
This research has made use of the Unified Cluster Catalogue (UCC), 
the SIMBAD database, operated at CDS, Strasbourg, France and the WEBDA database, 
operated at the Department of Theoretical
Physics and Astrophysics of Masaryk University.
This work has made use of data from the European Space Agency (ESA) mission
{\it Gaia} (\url{https://www.cosmos.esa.int/gaia}), processed by the {\it Gaia}
Data Processing and Analysis Consortium (DPAC,
\url{https://www.cosmos.esa.int/web/gaia/dpac/consortium}). Funding for the DPAC
has been provided by national institutions, in particular the institutions
participating in the {\it Gaia} Multilateral Agreement.

\end{acknowledgements}

\bibliographystyle{aa} 
\bibliography{paper.bib}

@ARTICLE{2024A&A...686A..42H,
       author = {{Hunt}, Emily L. and {Reffert}, Sabine},
        title = "{Improving the open cluster census. III. Using cluster masses, radii, and dynamics to create a cleaned open cluster catalogue}",
      journal = {\aap},
         year = 2024,
        month = jun,
       volume = {686},
          eid = {A42},
        pages = {A42},
          doi = {10.1051/0004-6361/202348662},
archivePrefix = {arXiv},
       eprint = {2403.05143},
 primaryClass = {astro-ph.GA},
       adsurl = {https://ui.adsabs.harvard.edu/abs/2024A&A...686A..42H}
}

@ARTICLE{1975PASP...87..425G,
       author = {{Gutierrez-Moreno}, A. and {Moreno}, H.},
        title = "{The ratio of color excesses in UBV photometry}",
      journal = {\pasp},
         year = 1975,
        month = jun,
       volume = {87},
        pages = {425-432},
          doi = {10.1086/129787},
       adsurl = {https://ui.adsabs.harvard.edu/abs/1975PASP...87..425G}
}

@ARTICLE{2023MNRAS.526.4107P,
       author = {{Perren}, Gabriel I. and {Pera}, Mar{\'\i}a S. and {Navone}, Hugo D. and {V{\'a}zquez}, Rub{\'e}n A.},
        title = "{The Unified Cluster Catalogue: towards a comprehensive and homogeneous data base of stellar clusters}",
      journal = {\mnras},
         year = 2023,
        month = dec,
       volume = {526},
       number = {3},
        pages = {4107-4119},
          doi = {10.1093/mnras/stad2826},
       adsurl = {https://ui.adsabs.harvard.edu/abs/2023MNRAS.526.4107P}
}

@ARTICLE{2023A&A...673A.114H,
       author = {{Hunt}, Emily L. and {Reffert}, Sabine},
        title = "{Improving the open cluster census. II. An all-sky cluster catalogue with Gaia DR3}",
      journal = {\aap},
         year = 2023,
        month = may,
       volume = {673},
          eid = {A114},
        pages = {A114},
          doi = {10.1051/0004-6361/202346285},
archivePrefix = {arXiv},
       eprint = {2303.13424},
 primaryClass = {astro-ph.GA},
       adsurl = {https://ui.adsabs.harvard.edu/abs/2023A&A...673A.114H}
}

@ARTICLE{2021A&A...650A.109P,
       author = {{Pera}, M.~S. and {Perren}, G.~I. and {Moitinho}, A. and {Navone}, H.~D. and {Vazquez}, R.~A.},
        title = "{pyUPMASK: an improved unsupervised clustering algorithm}",
      journal = {\aap},
         year = 2021,
        month = jun,
       volume = {650},
          eid = {A109},
        pages = {A109},
          doi = {10.1051/0004-6361/202040252},
archivePrefix = {arXiv},
       eprint = {2101.01660},
 primaryClass = {astro-ph.GA},
       adsurl = {https://ui.adsabs.harvard.edu/abs/2021A&A...650A.109P}
}

@ARTICLE{2018ApJ...863...91F,
       author = {{Faherty}, Jacqueline K. and {Bochanski}, John J. and {Gagn{\'e}}, Jonathan and {Nelson}, Olivia and {Coker}, Kristina and {Smithka}, Iliya and {Desir}, Deion and {Vasquez}, Chelsea},
        title = "{New and Known Moving Groups and Clusters Identified in a Gaia Comoving Catalog}",
      journal = {\apj},
         year = 2018,
        month = aug,
       volume = {863},
       number = {1},
          eid = {91},
        pages = {91},
          doi = {10.3847/1538-4357/aac76e},
archivePrefix = {arXiv},
       eprint = {1804.09058},
 primaryClass = {astro-ph.SR},
       adsurl = {https://ui.adsabs.harvard.edu/abs/2018ApJ...863...91F}
}

@ARTICLE{1999PASP..111...63F,
       author = {{Fitzpatrick}, Edward L.},
        title = "{Correcting for the Effects of Interstellar Extinction}",
      journal = {\pasp},
         year = 1999,
        month = jan,
       volume = {111},
       number = {755},
        pages = {63-75},
          doi = {10.1086/316293},
archivePrefix = {arXiv},
       eprint = {astro-ph/9809387},
 primaryClass = {astro-ph},
       adsurl = {https://ui.adsabs.harvard.edu/abs/1999PASP..111...63F}
}

@ARTICLE{2023A&A...674A..33G,
       author = {{Gaia Collaboration} and {Montegriffo}, P. and {Bellazzini}, M. and {De Angeli}, F. and {Andrae}, R. and {Barstow}, M.~A. and {Bossini}, D. and {Bragaglia}, A. and {Burgess}, P.~W. and {Cacciari}, C. and {Carrasco}, J.~M. and {Chornay}, N. and {Delchambre}, L. and {Evans}, D.~W. and {Fouesneau}, M. and {Fr{\'e}mat}, Y. and {Garabato}, D. and {Jordi}, C. and {Manteiga}, M. and {Massari}, D. and {Palaversa}, L. and {Pancino}, E. and {Riello}, M. and {Ruz Mieres}, D. and {Sanna}, N. and {Santove{\~n}a}, R. and {Sordo}, R. and {Vallenari}, A. and {Walton}, N.~A. and {Brown}, A.~G.~A. and {Prusti}, T. and {de Bruijne}, J.~H.~J. and {Arenou}, F. and {Babusiaux}, C. and {Biermann}, M. and {Creevey}, O.~L. and {Ducourant}, C. and {Eyer}, L. and {Guerra}, R. and {Hutton}, A. and {Klioner}, S.~A. and {Lammers}, U.~L. and {Lindegren}, L. and {Luri}, X. and {Mignard}, F. and {Panem}, C. and {Pourbaix}, D. and {Randich}, S. and {Sartoretti}, P. and {Soubiran}, C. and {Tanga}, P. and {Bailer-Jones}, C.~A.~L. and {Bastian}, U. and {Drimmel}, R. and {Jansen}, F. and {Katz}, D. and {Lattanzi}, M.~G. and {van Leeuwen}, F. and {Bakker}, J. and {Casta{\~n}eda}, J. and {Fabricius}, C. and {Galluccio}, L. and {Guerrier}, A. and {Heiter}, U. and {Masana}, E. and {Messineo}, R. and {Mowlavi}, N. and {Nicolas}, C. and {Nienartowicz}, K. and {Pailler}, F. and {Panuzzo}, P. and {Riclet}, F. and {Roux}, W. and {Seabroke}, G.~M. and {Th{\'e}venin}, F. and {Gracia-Abril}, G. and {Portell}, J. and {Teyssier}, D. and {Altmann}, M. and {Audard}, M. and {Bellas-Velidis}, I. and {Benson}, K. and {Berthier}, J. and {Blomme}, R. and {Busonero}, D. and {Busso}, G. and {C{\'a}novas}, H. and {Carry}, B. and {Cellino}, A. and {Cheek}, N. and {Clementini}, G. and {Damerdji}, Y. and {Davidson}, M. and {de Teodoro}, P. and {Nu{\~n}ez Campos}, M. and {Dell'Oro}, A. and {Esquej}, P. and {Fern{\'a}ndez-Hern{\'a}ndez}, J. and {Fraile}, E. and {Garc{\'\i}a-Lario}, P. and {Gosset}, E. and {Haigron}, R. and {Halbwachs}, J. -L. and {Hambly}, N.~C. and {Harrison}, D.~L. and {Hern{\'a}ndez}, J. and {Hestroffer}, D. and {Hodgkin}, S.~T. and {Holl}, B. and {Jan{\ss}en}, K. and {Jevardat de Fombelle}, G. and {Jordan}, S. and {Krone-Martins}, A. and {Lanzafame}, A.~C. and {L{\"o}ffler}, W. and {Marchal}, O. and {Marrese}, P.~M. and {Moitinho}, A. and {Muinonen}, K. and {Osborne}, P. and {Pauwels}, T. and {Recio-Blanco}, A. and {Reyl{\'e}}, C. and {Rimoldini}, L. and {Roegiers}, T. and {Rybizki}, J. and {Sarro}, L.~M. and {Siopis}, C. and {Smith}, M. and {Sozzetti}, A. and {Utrilla}, E. and {van Leeuwen}, M. and {Abbas}, U. and {{\'A}brah{\'a}m}, P. and {Abreu Aramburu}, A. and {Aerts}, C. and {Aguado}, J.~J. and {Ajaj}, M. and {Aldea-Montero}, F. and {Altavilla}, G. and {{\'A}lvarez}, M.~A. and {Alves}, J. and {Anderson}, R.~I. and {Anglada Varela}, E. and {Antoja}, T. and {Baines}, D. and {Baker}, S.~G. and {Balaguer-N{\'u}{\~n}ez}, L. and {Balbinot}, E. and {Balog}, Z. and {Barache}, C. and {Barbato}, D. and {Barros}, M. and {Bartolom{\'e}}, S. and {Bassilana}, J. -L. and {Bauchet}, N. and {Becciani}, U. and {Berihuete}, A. and {Bernet}, M. and {Bertone}, S. and {Bianchi}, L. and {Binnenfeld}, A. and {Blanco-Cuaresma}, S. and {Boch}, T. and {Bombrun}, A. and {Bouquillon}, S. and {Bramante}, L. and {Breedt}, E. and {Bressan}, A. and {Brouillet}, N. and {Brugaletta}, E. and {Bucciarelli}, B. and {Burlacu}, A. and {Butkevich}, A.~G. and {Buzzi}, R. and {Caffau}, E. and {Cancelliere}, R. and {Cantat-Gaudin}, T. and {Carballo}, R. and {Carlucci}, T. and {Carnerero}, M.~I. and {Casamiquela}, L. and {Castellani}, M. and {Castro-Ginard}, A. and {Chaoul}, L. and {Charlot}, P. and {Chemin}, L. and {Chiaramida}, V. and {Chiavassa}, A. and {Comoretto}, G. and {Contursi}, G. and {Cooper}, W.~J. and {Cornez}, T. and {Cowell}, S. and {Crifo}, F. and {Cropper}, M. and {Crosta}, M. and {Crowley}, C. and {Dafonte}, C. and {Dapergolas}, A. and {David}, P. and {de Laverny}, P. and {De Luise}, F. and {De March}, R. and {De Ridder}, J. and {de Souza}, R. and {de Torres}, A. and {del Peloso}, E.~F. and {del Pozo}, E. and {Delbo}, M. and {Delgado}, A. and {Delisle}, J. -B. and {Demouchy}, C. and {Dharmawardena}, T.~E. and {Diakite}, S. and {Diener}, C. and {Distefano}, E. and {Dolding}, C. and {Enke}, H. and {Fabre}, C. and {Fabrizio}, M. and {Faigler}, S. and {Fedorets}, G. and {Fernique}, P. and {Figueras}, F. and {Fournier}, Y. and {Fouron}, C. and {Fragkoudi}, F. and {Gai}, M. and {Garcia-Gutierrez}, A. and {Garcia-Reinaldos}, M. and {Garc{\'\i}a-Torres}, M. and {Garofalo}, A. and {Gavel}, A. and {Gavras}, P. and {Gerlach}, E. and {Geyer}, R. and {Giacobbe}, P. and {Gilmore}, G. and {Girona}, S. and {Giuffrida}, G. and {Gomel}, R. and {Gomez}, A. and {Gonz{\'a}lez-N{\'u}{\~n}ez}, J. and {Gonz{\'a}lez-Santamar{\'\i}a}, I. and {Gonz{\'a}lez-Vidal}, J.~J. and {Granvik}, M. and {Guillout}, P. and {Guiraud}, J. and {Guti{\'e}rrez-S{\'a}nchez}, R. and {Guy}, L.~P. and {Hatzidimitriou}, D. and {Hauser}, M. and {Haywood}, M. and {Helmer}, A. and {Helmi}, A. and {Sarmiento}, M.~H. and {Hidalgo}, S.~L. and {H{\l}adczuk}, N. and {Hobbs}, D. and {Holland}, G. and {Huckle}, H.~E. and {Jardine}, K. and {Jasniewicz}, G. and {Jean-Antoine Piccolo}, A. and {Jim{\'e}nez-Arranz}, {\'O}. and {Juaristi Campillo}, J. and {Julbe}, F. and {Karbevska}, L. and {Kervella}, P. and {Khanna}, S. and {Kordopatis}, G. and {Korn}, A.~J. and {K{\'o}sp{\'a}l}, {\'A}. and {Kostrzewa-Rutkowska}, Z. and {Kruszy{\'n}ska}, K. and {Kun}, M. and {Laizeau}, P. and {Lambert}, S. and {Lanza}, A.~F. and {Lasne}, Y. and {Le Campion}, J. -F. and {Lebreton}, Y. and {Lebzelter}, T. and {Leccia}, S. and {Leclerc}, N. and {Lecoeur-Taibi}, I. and {Liao}, S. and {Licata}, E.~L. and {Lindstr{\'o}m}, H.~E.~P. and {Lister}, T.~A. and {Livanou}, E. and {Lobel}, A. and {Lorca}, A. and {Loup}, C. and {Madrero Pardo}, P. and {Magdaleno Romeo}, A. and {Managau}, S. and {Mann}, R.~G. and {Marchant}, J.~M. and {Marconi}, M. and {Marcos}, J. and {Marcos Santos}, M.~M.~S. and {Mar{\'\i}n Pina}, D. and {Marinoni}, S. and {Marocco}, F. and {Marshall}, D.~J. and {Martin Polo}, L. and {Mart{\'\i}n-Fleitas}, J.~M. and {Marton}, G. and {Mary}, N. and {Masip}, A. and {Mastrobuono-Battisti}, A. and {Mazeh}, T. and {McMillan}, P.~J. and {Messina}, S. and {Michalik}, D. and {Millar}, N.~R. and {Mints}, A. and {Molina}, D. and {Molinaro}, R. and {Moln{\'a}r}, L. and {Monari}, G. and {Mongui{\'o}}, M. and {Montero}, A. and {Mor}, R. and {Mora}, A. and {Morbidelli}, R. and {Morel}, T. and {Morris}, D. and {Muraveva}, T. and {Murphy}, C.~P. and {Musella}, I. and {Nagy}, Z. and {Noval}, L. and {Oca{\~n}a}, F. and {Ogden}, A. and {Ordenovic}, C. and {Osinde}, J.~O. and {Pagani}, C. and {Pagano}, I. and {Palicio}, P.~A. and {Pallas-Quintela}, L. and {Panahi}, A. and {Payne-Wardenaar}, S. and {Pe{\~n}alosa Esteller}, X. and {Penttil{\"a}}, A. and {Pichon}, B. and {Piersimoni}, A.~M. and {Pineau}, F. -X. and {Plachy}, E. and {Plum}, G. and {Poggio}, E. and {Pr{\v{s}}a}, A. and {Pulone}, L. and {Racero}, E. and {Ragaini}, S. and {Rainer}, M. and {Raiteri}, C.~M. and {Ramos}, P. and {Ramos-Lerate}, M. and {Re Fiorentin}, P. and {Regibo}, S. and {Richards}, P.~J. and {Rios Diaz}, C. and {Ripepi}, V. and {Riva}, A. and {Rix}, H. -W. and {Rixon}, G. and {Robichon}, N. and {Robin}, A.~C. and {Robin}, C. and {Roelens}, M. and {Rogues}, H.~R.~O. and {Rohrbasser}, L. and {Romero-G{\'o}mez}, M. and {Rowell}, N. and {Royer}, F. and {Rybicki}, K.~A. and {Sadowski}, G. and {S{\'a}ez N{\'u}{\~n}ez}, A. and {Sagrist{\`a} Sell{\'e}s}, A. and {Sahlmann}, J. and {Salguero}, E. and {Samaras}, N. and {Sanchez Gimenez}, V. and {Sarasso}, M. and {Schultheis}, M.~S. and {Sciacca}, E. and {Segol}, M. and {Segovia}, J.~C. and {S{\'e}gransan}, D. and {Semeux}, D. and {Shahaf}, S. and {Siddiqui}, H.~I. and {Siebert}, A. and {Siltala}, L. and {Silvelo}, A. and {Slezak}, E. and {Slezak}, I. and {Smart}, R.~L. and {Snaith}, O.~N. and {Solano}, E. and {Solitro}, F. and {Souami}, D. and {Souchay}, J. and {Spagna}, A. and {Spina}, L. and {Spoto}, F. and {Steele}, I.~A. and {Steidelm{\"u}ller}, H. and {Stephenson}, C.~A. and {S{\"u}veges}, M. and {Surdej}, J. and {Szabados}, L. and {Szegedi-Elek}, E. and {Taris}, F. and {Taylor}, M.~B. and {Teixeira}, R. and {Tolomei}, L. and {Tonello}, N. and {Torra}, F. and {Torra}, J. and {Torralba Elipe}, G. and {Trabucchi}, M. and {Tsounis}, A.~T. and {Turon}, C. and {Ulla}, A. and {Unger}, N. and {Vaillant}, M.~V. and {van Dillen}, E. and {van Reeven}, W. and {Vanel}, O. and {Vecchiato}, A. and {Viala}, Y. and {Vicente}, D. and {Voutsinas}, S. and {Wevers}, T. and {Wyrzykowski}, {\L}. and {Yoldas}, A. and {Yvard}, P. and {Zhao}, H. and {Zorec}, J. and {Zucker}, S. and {Zwitter}, T.},
        title = "{Gaia Data Release 3. The Galaxy in your preferred colours: Synthetic photometry from Gaia low-resolution spectra}",
      journal = {\aap},
         year = 2023,
        month = jun,
       volume = {674},
          eid = {A33},
        pages = {A33},
          doi = {10.1051/0004-6361/202243709},
archivePrefix = {arXiv},
       eprint = {2206.06215},
 primaryClass = {astro-ph.SR},
       adsurl = {https://ui.adsabs.harvard.edu/abs/2023A&A...674A..33G}
}

@ARTICLE{2023A&A...674A...3M,
       author = {{Montegriffo}, P. and {De Angeli}, F. and {Andrae}, R. and {Riello}, M. and {Pancino}, E. and {Sanna}, N. and {Bellazzini}, M. and {Evans}, D.~W. and {Carrasco}, J.~M. and {Sordo}, R. and {Busso}, G. and {Cacciari}, C. and {Jordi}, C. and {van Leeuwen}, F. and {Vallenari}, A. and {Altavilla}, G. and {Barstow}, M.~A. and {Brown}, A.~G.~A. and {Burgess}, P.~W. and {Castellani}, M. and {Cowell}, S. and {Davidson}, M. and {De Luise}, F. and {Delchambre}, L. and {Diener}, C. and {Fabricius}, C. and {Fr{\'e}mat}, Y. and {Fouesneau}, M. and {Gilmore}, G. and {Giuffrida}, G. and {Hambly}, N.~C. and {Harrison}, D.~L. and {Hidalgo}, S. and {Hodgkin}, S.~T. and {Holland}, G. and {Marinoni}, S. and {Osborne}, P.~J. and {Pagani}, C. and {Palaversa}, L. and {Piersimoni}, A.~M. and {Pulone}, L. and {Ragaini}, S. and {Rainer}, M. and {Richards}, P.~J. and {Rowell}, N. and {Ruz-Mieres}, D. and {Sarro}, L.~M. and {Walton}, N.~A. and {Yoldas}, A.},
        title = "{Gaia Data Release 3. External calibration of BP/RP low-resolution spectroscopic data}",
      journal = {\aap},
         year = 2023,
        month = jun,
       volume = {674},
          eid = {A3},
        pages = {A3},
          doi = {10.1051/0004-6361/202243880},
archivePrefix = {arXiv},
       eprint = {2206.06205},
 primaryClass = {astro-ph.IM},
       adsurl = {https://ui.adsabs.harvard.edu/abs/2023A&A...674A...3M}
}

@ARTICLE{1997A&AS..124..349M,
   author = {{Mermilliod}, J.-C. and {Mermilliod}, M. and {Hauck}, B.},
    title = "{The General Catalogue of Photometric Data (GCPD). II.}",
  journal = {\aaps},
     year = 1997,
    month = aug,
   volume = 124,
      doi = {10.1051/aas:1997197},
   adsurl = {http://adsabs.harvard.edu/abs/1997A%26AS..124..349M}
}

@ARTICLE{1978A&A....62..411B,
       author = {{Buser}, R.},
        title = "{A systematic investigation of multicolor photometric systems. I. The UBV, RGU and uvby systems.}",
      journal = {\aap},
         year = 1978,
        month = jan,
       volume = {62},
        pages = {411-424},
       adsurl = {https://ui.adsabs.harvard.edu/abs/1978A&A....62..411B}
}

@ARTICLE{1958LowOB...4...37J,
       author = {{Johnson}, Harold Lester},
        title = "{The intrinsic colors of early-type stars}",
      journal = {Lowell Observatory Bulletin},
         year = 1958,
        month = jan,
       volume = {4},
       number = {90},
        pages = {37-46},
       adsurl = {https://ui.adsabs.harvard.edu/abs/1958LowOB...4...37J}
}

@ARTICLE{2023AJ....165...79Y,
       author = {{Yontan}, Talar},
        title = "{An Investigation of Open Clusters Berkeley 68 and Stock 20 Using CCD UBV and Gaia DR3 Data}",
      journal = {\aj},
         year = 2023,
        month = mar,
       volume = {165},
       number = {3},
          eid = {79},
        pages = {79},
          doi = {10.3847/1538-3881/aca6f0},
archivePrefix = {arXiv},
       eprint = {2211.09825},
 primaryClass = {astro-ph.GA},
       adsurl = {https://ui.adsabs.harvard.edu/abs/2023AJ....165...79Y}
}

@ARTICLE{2024ApJS..272...20A,
       author = {{An}, Deokkeun and {Beers}, Timothy C. and {Chiti}, Anirudh},
        title = "{A Blueprint for the Milky Way's Stellar Populations. V. 3D Local Dust Extinction}",
      journal = {\apjs},
         year = 2024,
        month = may,
       volume = {272},
       number = {1},
          eid = {20},
        pages = {20},
          doi = {10.3847/1538-4365/ad3641},
archivePrefix = {arXiv},
       eprint = {2404.14626},
 primaryClass = {astro-ph.GA},
       adsurl = {https://ui.adsabs.harvard.edu/abs/2024ApJS..272...20A}
}

@ARTICLE{2025A&A...701A.258N,
       author = {{Nguyen}, C.~T. and {Costa}, G. and {Bressan}, A. and {Girardi}, L. and {Cescutti}, G. and {Korn}, A.~J. and {Volpato}, G. and {Chen}, Y. and {Pastorelli}, G. and {Trabucchi}, M. and {Shepherd}, K.~G. and {Ettorre}, G. and {Zaggia}, S.},
        title = "{PARSEC V2.0: Rotating tracks and isochrones for seven additional metallicities in the range Z = 0.0001{\textendash}0.03}",
      journal = {\aap},
         year = 2025,
        month = sep,
       volume = {701},
          eid = {A258},
        pages = {A258},
          doi = {10.1051/0004-6361/202556005},
archivePrefix = {arXiv},
       eprint = {2508.02393},
 primaryClass = {astro-ph.SR},
       adsurl = {https://ui.adsabs.harvard.edu/abs/2025A&A...701A.258N}
}

@ARTICLE{2022MNRAS.509..421N,
       author = {{Netopil}, Martin and {Oralhan}, {\.I}nci Akkaya and {{\c{C}}akmak}, Hikmet and {Michel}, Ra{\'u}l and {Karata{\c{s}}}, Y{\"u}ksel},
        title = "{The Galactic metallicity gradient shown by open clusters in the light of radial migration}",
      journal = {\mnras},
         year = 2022,
        month = jan,
       volume = {509},
       number = {1},
        pages = {421-439},
          doi = {10.1093/mnras/stab2961},
archivePrefix = {arXiv},
       eprint = {2110.11893},
 primaryClass = {astro-ph.GA},
       adsurl = {https://ui.adsabs.harvard.edu/abs/2022MNRAS.509..421N}
}

@ARTICLE{2022A&A...664A.174V,
       author = {{Vergely}, J.~L. and {Lallement}, R. and {Cox}, N.~L.~J.},
        title = "{Three-dimensional extinction maps: Inverting inter-calibrated extinction catalogues}",
      journal = {\aap},
         year = 2022,
        month = aug,
       volume = {664},
          eid = {A174},
        pages = {A174},
          doi = {10.1051/0004-6361/202243319},
archivePrefix = {arXiv},
       eprint = {2205.09087},
 primaryClass = {astro-ph.GA},
       adsurl = {https://ui.adsabs.harvard.edu/abs/2022A&A...664A.174V}
}

@ARTICLE{2001MNRAS.328..370Y,
       author = {{Yadav}, R.~K.~S. and {Sagar}, Ram},
        title = "{Non-uniform extinction in young open star clusters}",
      journal = {\mnras},
         year = 2001,
        month = dec,
       volume = {328},
       number = {2},
        pages = {370-380},
          doi = {10.1046/j.1365-8711.2001.04754.x},
archivePrefix = {arXiv},
       eprint = {astro-ph/0106260},
 primaryClass = {astro-ph},
       adsurl = {https://ui.adsabs.harvard.edu/abs/2001MNRAS.328..370Y}
}

@ARTICLE{2021MNRAS.505.5957B,
       author = {{Baumgardt}, H. and {Vasiliev}, E.},
        title = "{Accurate distances to Galactic globular clusters through a combination of Gaia EDR3, HST, and literature data}",
      journal = {\mnras},
         year = 2021,
        month = aug,
       volume = {505},
       number = {4},
        pages = {5957-5977},
          doi = {10.1093/mnras/stab1474},
archivePrefix = {arXiv},
       eprint = {2105.09526},
 primaryClass = {astro-ph.GA},
       adsurl = {https://ui.adsabs.harvard.edu/abs/2021MNRAS.505.5957B}
}

@ARTICLE{2010arXiv1012.3224H,
       author = {{Harris}, William E.},
        title = "{A New Catalog of Globular Clusters in the Milky Way}",
      journal = {arXiv e-prints},
         year = 2010,
        month = dec,
          eid = {arXiv:1012.3224},
        pages = {arXiv:1012.3224},
          doi = {10.48550/arXiv.1012.3224},
archivePrefix = {arXiv},
       eprint = {1012.3224},
 primaryClass = {astro-ph.GA},
       adsurl = {https://ui.adsabs.harvard.edu/abs/2010arXiv1012.3224H}
}

@ARTICLE{2001ApJ...562..852H,
       author = {{Hartmann}, Lee and {Ballesteros-Paredes}, Javier and {Bergin}, Edwin A.},
        title = "{Rapid Formation of Molecular Clouds and Stars in the Solar Neighborhood}",
      journal = {\apj},
         year = 2001,
        month = dec,
       volume = {562},
       number = {2},
        pages = {852-868},
          doi = {10.1086/323863},
archivePrefix = {arXiv},
       eprint = {astro-ph/0108023},
 primaryClass = {astro-ph},
       adsurl = {https://ui.adsabs.harvard.edu/abs/2001ApJ...562..852H}
}

@ARTICLE{2023ApJ...946...43W,
       author = {{Wang}, Shu and {Chen}, Xiaodian},
        title = "{The Optical to Infrared Extinction Law of Magellanic Clouds Based on Red Supergiants and Classical Cepheids}",
      journal = {\apj},
         year = 2023,
        month = mar,
       volume = {946},
       number = {1},
          eid = {43},
        pages = {43},
          doi = {10.3847/1538-4357/acb647},
archivePrefix = {arXiv},
       eprint = {2301.09146},
 primaryClass = {astro-ph.GA},
       adsurl = {https://ui.adsabs.harvard.edu/abs/2023ApJ...946...43W}
}

@ARTICLE{2015A&A...582A..19N,
       author = {{Netopil}, M. and {Paunzen}, E. and {Carraro}, G.},
        title = "{A comparative study on the reliability of open cluster parameters}",
      journal = {\aap},
         year = 2015,
        month = oct,
       volume = {582},
          eid = {A19},
        pages = {A19},
          doi = {10.1051/0004-6361/201526372},
archivePrefix = {arXiv},
       eprint = {1508.01296},
 primaryClass = {astro-ph.SR},
       adsurl = {https://ui.adsabs.harvard.edu/abs/2015A&A...582A..19N}
}

@ARTICLE{2025AJ....170..288L,
       author = {{Li}, Lu and {Shao}, Zhengyi and {Li}, Zhaozhou and {Fu}, Xiaoting},
        title = "{The MiMO Catalog: Physical Parameters and Stellar Mass Functions of 1232 Open Clusters from Gaia DR3}",
      journal = {\aj},
         year = 2025,
        month = nov,
       volume = {170},
       number = {5},
          eid = {288},
        pages = {288},
          doi = {10.3847/1538-3881/ae0cb6},
archivePrefix = {arXiv},
       eprint = {2510.23374},
 primaryClass = {astro-ph.GA},
       adsurl = {https://ui.adsabs.harvard.edu/abs/2025AJ....170..288L}
}

@ARTICLE{2025A&A...694A..70M,
       author = {{Moreira}, Sandro and {Moitinho}, Andr{\'e} and {Silva}, Andr{\'e} and {Almeida}, Duarte},
        title = "{Modelling the evolution of the Galactic disc scale height traced by open clusters}",
      journal = {\aap},
         year = 2025,
        month = feb,
       volume = {694},
          eid = {A70},
        pages = {A70},
          doi = {10.1051/0004-6361/202450369},
archivePrefix = {arXiv},
       eprint = {2406.14661},
 primaryClass = {astro-ph.GA},
       adsurl = {https://ui.adsabs.harvard.edu/abs/2025A&A...694A..70M}
}

@ARTICLE{2020MNRAS.499.1874M,
       author = {{Monteiro}, H. and {Dias}, W.~S. and {Moitinho}, A. and {Cantat-Gaudin}, T. and {L{\'e}pine}, J.~R.~D. and {Carraro}, G. and {Paunzen}, E.},
        title = "{Fundamental parameters for 45 open clusters with Gaia DR2, an improved extinction correction and a metallicity gradient prior}",
      journal = {\mnras},
         year = 2020,
        month = dec,
       volume = {499},
       number = {2},
        pages = {1874-1889},
          doi = {10.1093/mnras/staa2983},
archivePrefix = {arXiv},
       eprint = {2009.11834},
 primaryClass = {astro-ph.SR},
       adsurl = {https://ui.adsabs.harvard.edu/abs/2020MNRAS.499.1874M}
}

@ARTICLE{2004AJ....128.2144M,
       author = {{McCall}, Marshall L.},
        title = "{On Determining Extinction from Reddening}",
      journal = {\aj},
         year = 2004,
        month = nov,
       volume = {128},
       number = {5},
        pages = {2144-2169},
          doi = {10.1086/424933},
       adsurl = {https://ui.adsabs.harvard.edu/abs/2004AJ....128.2144M}
}

@ARTICLE{2024ApJ...973..136O,
       author = {{O'Neill}, Theo J. and {Zucker}, Catherine and {Goodman}, Alyssa A. and {Edenhofer}, Gordian},
        title = "{The Local Bubble Is a Local Chimney: A New Model from 3D Dust Mapping}",
      journal = {\apj},
         year = 2024,
        month = oct,
       volume = {973},
       number = {2},
          eid = {136},
        pages = {136},
          doi = {10.3847/1538-4357/ad61de},
archivePrefix = {arXiv},
       eprint = {2403.04961},
 primaryClass = {astro-ph.GA},
       adsurl = {https://ui.adsabs.harvard.edu/abs/2024ApJ...973..136O}
}

@ARTICLE{2020ARA&A..58..157T,
       author = {{Tacconi}, Linda J. and {Genzel}, Reinhard and {Sternberg}, Amiel},
        title = "{The Evolution of the Star-Forming Interstellar Medium Across Cosmic Time}",
      journal = {\araa},
         year = 2020,
        month = aug,
       volume = {58},
        pages = {157-203},
          doi = {10.1146/annurev-astro-082812-141034},
archivePrefix = {arXiv},
       eprint = {2003.06245},
 primaryClass = {astro-ph.GA},
       adsurl = {https://ui.adsabs.harvard.edu/abs/2020ARA&A..58..157T}
}

@ARTICLE{2001ApJ...548..296W,
       author = {{Weingartner}, Joseph C. and {Draine}, B.~T.},
        title = "{Dust Grain-Size Distributions and Extinction in the Milky Way, Large Magellanic Cloud, and Small Magellanic Cloud}",
      journal = {\apj},
         year = 2001,
        month = feb,
       volume = {548},
       number = {1},
        pages = {296-309},
          doi = {10.1086/318651},
archivePrefix = {arXiv},
       eprint = {astro-ph/0008146},
 primaryClass = {astro-ph},
       adsurl = {https://ui.adsabs.harvard.edu/abs/2001ApJ...548..296W}
}

@ARTICLE{2003ARA&A..41..241D,
       author = {{Draine}, B.~T.},
        title = "{Interstellar Dust Grains}",
      journal = {\araa},
         year = 2003,
        month = jan,
       volume = {41},
        pages = {241-289},
          doi = {10.1146/annurev.astro.41.011802.094840},
archivePrefix = {arXiv},
       eprint = {astro-ph/0304489},
 primaryClass = {astro-ph},
       adsurl = {https://ui.adsabs.harvard.edu/abs/2003ARA&A..41..241D}
}

\begin{appendix} 

\section{Table of results}
Table \ref{table_master} lists essential data and results for our sample of star clusters. 
It is organised as follows:

\begin{itemize}
\item Column 1: Name.
\item Column 2: Distance from the Sun taken from the literature.
\item Column 3: Absorption in the $V$ band taken from the literature.
\item Column 4: Galactic longitude (J2000).
\item Column 5: Galactic latitude (J2000).
\item Column 6: X, in the direction of the Galactic centre.
\item Column 7: Y, in the direction of the Galactic disk rotation.
\item Column 8: Z, in the direction of the North Galactic Pole.
\item Column 9: Galactocentric distance.
\item Column 10: $E(B-V)$.
\item Column 11: $\sigma E(B-V)$.
\item Column 12: $E(BP-RP)$.
\item Column 13: $\sigma E(BP-RP)$.
\item Column 14: $E(U-B)$.
\item Column 15: $\sigma E(U-B)$.
\item Column 16: $E(U-BP)$.
\item Column 17: $\sigma E(U-BP)$.
\item Column 18: $E(U-G)$.
\item Column 19: $\sigma E(U-G)$.
\item Column 20: $E(U-RP)$.
\item Column 21: $\sigma E(U-RP)$.
\item Column 22: Number of used members.
\end{itemize}

\setcounter{table}{0}  
\begin{table*}
\caption{Essential data and results for our sample of star clusters, sorted by increasing Galactic longitude.
The full table with all 5377 entries is only available online.}
\begin{center}
\begin{adjustbox}{max width=1.2\textwidth,angle=90}
\label{table_master}
\tiny
\begin{tabular}{cccccccccccccccccccccc}
\hline
(1) & (2) & (3) & (4) & (5) & (6) & (7) & (8) & (9) & (10) & (11) & (12) & (13) & (14) & (15) & (16) 
& (17) & (18) & (19) & (20) & (21) & (22)  \\ 
\hline
HSC 1	&	472	&	0.50	&	0.025	&	$-$21.310	&	440	&	0	&	$-$171	&	7728	&	0.098	&	0.003	&	0.195	&	0.017	&	0.085	&	0.020	&	0.150	&	0.035	&	0.195	&	0.005	&	0.335	&	0.055	&	12	\\
ESO 456 05	&	3592	&	5.68	&	0.054	&	$-$1.841	&	3590	&	3	&	$-$115	&	4608	&	0.478	&	0.014	&	0.749	&	0.027	&	0.303	&	0.003	&	0.633	&	0.018	&	0.978	&	0.023	&	1.353	&	0.008	&	12	\\
Theia 67	&	147	&	0.65	&	0.094	&	+6.891	&	146	&	0	&	18	&	8053	&	0.049	&	0.007	&	0.139	&	0.002	&	0.038	&	0.013	&	0.060	&	0.020	&	0.118	&	0.033	&	0.205	&	0.025	&	7	\\
UBC 1001	&	2769	&	1.91	&	0.168	&	$-$0.866	&	2768	&	8	&	$-$42	&	5431	&	0.673	&	0.008	&	1.035	&	0.102	&	0.358	&	0.013	&	0.835	&	0.005	&	1.525	&	0.245	&	1.810	&	0.005	&	21	\\
Casado 7	&	1376	&	2.30	&	0.214	&	$-$0.509	&	1376	&	5	&	$-$12	&	6824	&	0.566	&	0.007	&	0.850	&	0.009	&	0.383	&	0.003	&	0.770	&	0.010	&	1.133	&	0.003	&	1.598	&	0.008	&	17	\\
UBC 1593	&	1753	&	1.65	&	0.214	&	$-$0.507	&	1752	&	7	&	$-$15	&	6447	&	0.705	&	0.023	&	0.564	&	0.008	&	0.445	&	0.025	&	0.825	&	0.091	&	1.200	&	0.205	&	1.645	&	0.380	&	7	\\
Ruprecht 134	&	2365	&	2.50	&	0.285	&	$-$1.639	&	2364	&	12	&	$-$68	&	5835	&	0.303	&	0.015	&	0.535	&	0.019	&	0.370	&	0.025	&	0.578	&	0.028	&	0.773	&	0.048	&	1.143	&	0.038	&	77	\\
UFMG 88	&	1261	&	2.29	&	0.342	&	$-$1.639	&	1260	&	8	&	$-$36	&	6939	&	0.435	&	0.011	&	0.656	&	0.015	&	0.310	&	0.000	&	0.603	&	0.003	&	0.875	&	0.015	&	1.268	&	0.003	&	26	\\
HSC 2	&	1562	&	1.68	&	0.459	&	$-$1.935	&	1561	&	12	&	$-$53	&	6638	&	0.340	&	0.004	&	0.554	&	0.002	&	0.120	&	0.000	&	0.330	&	0.010	&	0.525	&	0.005	&	0.910	&	0.035	&	17	\\
HSC 3	&	1436	&	1.08	&	0.557	&	$-$6.955	&	1425	&	14	&	$-$174	&	6764	&	0.408	&	0.009	&	0.650	&	0.000	&	0.300	&	0.010	&	0.568	&	0.008	&	0.830	&	0.005	&	1.203	&	0.023	&	15	\\
UBC 91	&	2272	&	1.37	&	0.616	&	$-$0.674	&	2271	&	24	&	$-$27	&	5928	&	0.471	&	0.013	&	0.731	&	0.015	&	0.378	&	0.003	&	0.703	&	0.008	&	1.015	&	0.010	&	1.398	&	0.058	&	7	\\
Ferrero 1	&	835	&	0.41	&	0.649	&	$-$8.140	&	827	&	9	&	$-$118	&	7365	&	0.135	&	0.006	&	0.271	&	0.011	&	0.105	&	0.000	&	0.210	&	0.020	&	0.318	&	0.028	&	0.440	&	0.055	&	60	\\
UFMG 92	&	2017	&	2.47	&	0.676	&	$-$2.251	&	2016	&	24	&	$-$79	&	6183	&	0.503	&	0.006	&	0.788	&	0.011	&	0.170	&	0.005	&	0.510	&	0.000	&	0.830	&	0.010	&	1.313	&	0.013	&	47	\\
HSC 6	&	3514	&	3.48	&	0.689	&	$-$0.920	&	3513	&	42	&	$-$56	&	4686	&	0.088	&	0.003	&	0.263	&	0.003	&	0.218	&	0.018	&	0.260	&	0.015	&	0.378	&	0.013	&	0.500	&	0.025	&	7	\\
HSC 8	&	2619	&	1.66	&	0.850	&	$-$0.812	&	2618	&	39	&	$-$37	&	5581	&	0.423	&	0.006	&	0.668	&	0.011	&	0.148	&	0.003	&	0.430	&	0.000	&	0.718	&	0.008	&	1.103	&	0.008	&	10	\\
HSC 9	&	1724	&	2.18	&	0.961	&	$-$2.249	&	1722	&	29	&	$-$68	&	6476	&	0.539	&	0.004	&	0.844	&	0.020	&	0.278	&	0.008	&	0.655	&	0.015	&	1.010	&	0.035	&	1.473	&	0.008	&	14	\\
UFMG 71	&	1266	&	2.81	&	0.967	&	+1.471	&	1265	&	21	&	33	&	6934	&	0.083	&	0.002	&	0.334	&	0.011	&	0.150	&	0.000	&	0.220	&	0.025	&	0.378	&	0.028	&	0.538	&	0.048	&	18	\\
HSC 12	&	871	&	0.65	&	1.115	&	$-$3.992	&	869	&	17	&	$-$61	&	7329	&	0.113	&	0.003	&	0.218	&	0.013	&	0.068	&	0.003	&	0.153	&	0.023	&	0.238	&	0.018	&	0.365	&	0.035	&	20	\\
HSC 13	&	2092	&	3.17	&	1.161	&	$-$1.997	&	2090	&	42	&	$-$73	&	6109	&	0.245	&	0.013	&	0.523	&	0.008	&	0.348	&	0.003	&	0.520	&	0.010	&	0.710	&	0.040	&	0.995	&	0.070	&	13	\\
UFMG 93	&	3006	&	2.92	&	1.188	&	$-$2.419	&	3003	&	62	&	$-$127	&	5195	&	0.760	&	0.017	&	0.501	&	0.007	&	0.250	&	0.075	&	0.623	&	0.258	&	0.960	&	0.395	&	1.386	&	0.506	&	14	\\
UFMG 83	&	2611	&	3.76	&	1.405	&	$-$0.899	&	2610	&	64	&	$-$41	&	5590	&	0.303	&	0.018	&	0.541	&	0.013	&	0.388	&	0.008	&	0.605	&	0.005	&	0.800	&	0.000	&	1.110	&	0.055	&	18	\\
UFMG 95	&	1007	&	0.74	&	1.521	&	$-$3.454	&	1004	&	27	&	$-$61	&	7194	&	0.111	&	0.004	&	0.208	&	0.004	&	0.063	&	0.008	&	0.143	&	0.003	&	0.223	&	0.013	&	0.340	&	0.005	&	30	\\
HSC 17	&	1262	&	1.91	&	1.652	&	$-$1.421	&	1261	&	36	&	$-$31	&	6939	&	0.161	&	0.022	&	0.463	&	0.004	&	0.295	&	0.005	&	0.408	&	0.018	&	0.558	&	0.068	&	0.760	&	0.130	&	11	\\
CWNU 2050	&	1534	&	1.10	&	1.861	&	$-$2.763	&	1531	&	50	&	$-$74	&	6667	&	0.238	&	0.003	&	0.400	&	0.000	&	0.168	&	0.008	&	0.310	&	0.000	&	0.440	&	0.000	&	0.693	&	0.018	&	13	\\
HSC 26	&	2259	&	3.51	&	2.296	&	$-$0.886	&	2257	&	91	&	$-$35	&	5943	&	0.270	&	0.018	&	0.579	&	0.007	&	0.388	&	0.003	&	0.595	&	0.020	&	0.810	&	0.065	&	1.113	&	0.088	&	17	\\
HSC 27	&	993	&	0.84	&	2.368	&	+0.538	&	992	&	41	&	9	&	7208	&	0.245	&	0.014	&	0.364	&	0.026	&	0.180	&	0.005	&	0.353	&	0.003	&	0.468	&	0.008	&	0.680	&	0.025	&	7	\\
HXHWL 29	&	565	&	0.90	&	2.387	&	$-$0.261	&	564	&	24	&	$-$3	&	7636	&	0.163	&	0.003	&	0.289	&	0.006	&	0.118	&	0.008	&	0.230	&	0.015	&	0.345	&	0.025	&	0.500	&	0.035	&	43	\\
CWNU 1494	&	1401	&	0.81	&	2.406	&	$-$2.777	&	1398	&	59	&	$-$68	&	6801	&	0.220	&	0.008	&	0.370	&	0.000	&	0.118	&	0.003	&	0.255	&	0.000	&	0.388	&	0.008	&	0.640	&	0.015	&	26	\\
UFMG 89	&	2214	&	4.15	&	2.438	&	$-$1.503	&	2211	&	94	&	$-$58	&	5989	&	0.510	&	0.026	&	1.430	&	0.000	&	0.401	&	0.061	&	1.018	&	0.283	&	1.479	&	0.414	&	2.100	&	0.630	&	9	\\
HSC 35	&	1481	&	3.38	&	2.639	&	+1.032	&	1479	&	68	&	27	&	6721	&	0.408	&	0.026	&	0.781	&	0.013	&	0.590	&	0.005	&	0.913	&	0.028	&	1.223	&	0.083	&	1.620	&	0.170	&	9	\\
HSC 34	&	1206	&	2.13	&	2.689	&	+1.369	&	1205	&	57	&	29	&	6995	&	0.586	&	0.011	&	0.901	&	0.016	&	0.340	&	0.010	&	0.770	&	0.010	&	1.165	&	0.015	&	1.660	&	0.025	&	14	\\
HSC 36	&	1890	&	4.05	&	2.723	&	$-$0.785	&	1888	&	90	&	$-$26	&	6312	&	0.969	&	0.011	&	1.453	&	0.040	&	0.870	&	0.005	&	1.640	&	0.015	&	2.335	&	0.110	&	3.078	&	0.023	&	7	\\
NGC 6520	&	1757	&	1.60	&	2.885	&	$-$2.848	&	1753	&	88	&	$-$87	&	6446	&	0.296	&	0.007	&	0.481	&	0.006	&	0.165	&	0.000	&	0.355	&	0.005	&	0.545	&	0.005	&	0.835	&	0.000	&	140	\\
HSC 38	&	2306	&	3.26	&	2.941	&	$-$1.239	&	2302	&	118	&	$-$50	&	5899	&	0.611	&	0.011	&	0.930	&	0.009	&	0.373	&	0.003	&	0.800	&	0.000	&	1.213	&	0.013	&	1.748	&	0.013	&	27	\\
NGC 6171	&	4764	&	1.34	&	3.387	&	+23.000	&	4378	&	259	&	1862	&	3455	&	0.519	&	0.013	&	0.846	&	0.042	&	0.313	&	0.008	&	0.670	&	0.005	&	1.050	&	0.010	&	1.533	&	0.008	&	9	\\
HSC 52	&	248	&	0.22	&	3.439	&	$-$57.771	&	132	&	8	&	$-$210	&	7952	&	0.039	&	0.004	&	0.048	&	0.004	&	0.000	&	0.000	&	0.000	&	0.000	&	0.018	&	0.013	&	0.060	&	0.030	&	12	\\
Theia 1661	&	2334	&	4.06	&	3.737	&	+1.080	&	2328	&	152	&	44	&	5873	&	0.366	&	0.029	&	0.760	&	0.025	&	0.440	&	0.000	&	0.750	&	0.030	&	0.998	&	0.088	&	1.385	&	0.185	&	52	\\
CWNU 1391	&	2173	&	2.81	&	3.766	&	$-$0.698	&	2169	&	143	&	$-$26	&	6033	&	0.220	&	0.033	&	0.574	&	0.021	&	0.358	&	0.008	&	0.553	&	0.038	&	0.728	&	0.088	&	1.010	&	0.175	&	9	\\
HSC 46	&	439	&	6.05	&	3.832	&	+46.791	&	300	&	20	&	320	&	7762	&	0.051	&	0.002	&	0.069	&	0.007	&	0.008	&	0.003	&	0.023	&	0.008	&	0.045	&	0.030	&	0.098	&	0.038	&	39	\\
Theia 2619	&	965	&	1.31	&	3.963	&	+0.667	&	963	&	67	&	11	&	7238	&	0.250	&	0.009	&	0.424	&	0.016	&	0.203	&	0.013	&	0.365	&	0.010	&	0.508	&	0.008	&	0.810	&	0.020	&	29	\\
Theia 5015	&	1057	&	1.90	&	4.385	&	+1.966	&	1054	&	81	&	36	&	7146	&	0.369	&	0.020	&	0.650	&	0.000	&	0.358	&	0.003	&	0.645	&	0.010	&	0.908	&	0.053	&	1.240	&	0.070	&	22	\\
UFMG 78	&	3201	&	3.32	&	4.472	&	+0.397	&	3191	&	250	&	22	&	5015	&	0.254	&	0.031	&	0.604	&	0.002	&	0.508	&	0.018	&	0.720	&	0.045	&	0.950	&	0.105	&	1.185	&	0.185	&	9	\\
Theia 70	&	212	&	0.68	&	4.519	&	+13.089	&	205	&	16	&	48	&	7989	&	0.115	&	0.009	&	0.209	&	0.013	&	0.093	&	0.003	&	0.158	&	0.028	&	0.268	&	0.008	&	0.373	&	0.043	&	10	\\
CWNU 1615	&	1271	&	2.06	&	4.583	&	$-$2.170	&	1266	&	102	&	$-$48	&	6933	&	0.449	&	0.032	&	0.841	&	0.067	&	0.253	&	0.013	&	0.655	&	0.070	&	0.985	&	0.090	&	1.365	&	0.240	&	27	\\
CWNU 488	&	3699	&	2.45	&	4.596	&	$-$0.548	&	3687	&	296	&	$-$35	&	4523	&	0.573	&	0.008	&	0.923	&	0.041	&	0.215	&	0.010	&	0.655	&	0.040	&	1.090	&	0.075	&	1.578	&	0.058	&	15	\\
Theia 1038	&	484	&	1.51	&	5.055	&	+0.853	&	482	&	43	&	7	&	7718	&	0.431	&	0.012	&	0.810	&	0.043	&	0.325	&	0.015	&	0.710	&	0.055	&	1.038	&	0.088	&	1.415	&	0.185	&	18	\\
UBC 94	&	1114	&	1.99	&	5.176	&	$-$0.281	&	1109	&	100	&	$-$5	&	7092	&	0.416	&	0.010	&	0.705	&	0.015	&	0.308	&	0.003	&	0.608	&	0.008	&	0.905	&	0.020	&	1.278	&	0.048	&	88	\\
Ruprecht 138	&	2131	&	2.78	&	5.304	&	$-$0.579	&	2122	&	197	&	$-$22	&	6081	&	0.130	&	0.015	&	0.450	&	0.020	&	0.210	&	0.005	&	0.338	&	0.043	&	0.468	&	0.063	&	0.718	&	0.133	&	43	\\
HSC 75	&	587	&	1.58	&	5.344	&	+2.776	&	584	&	55	&	28	&	7615	&	0.285	&	0.028	&	0.476	&	0.015	&	0.228	&	0.028	&	0.388	&	0.003	&	0.560	&	0.020	&	0.838	&	0.013	&	7	\\
HSC 76	&	1280	&	2.47	&	5.525	&	$-$0.568	&	1274	&	123	&	$-$13	&	6927	&	0.206	&	0.014	&	0.481	&	0.004	&	0.445	&	0.005	&	0.595	&	0.025	&	0.783	&	0.053	&	1.020	&	0.085	&	7	\\
\hline
\end{tabular}
\end{adjustbox}
\end{center}
\end{table*}

\section{Colour-colour and colour-magnitude diagrams} \label{ccds_appendix}

\begin{figure*}
    \centering
    \includegraphics[width = 0.6\textheight]{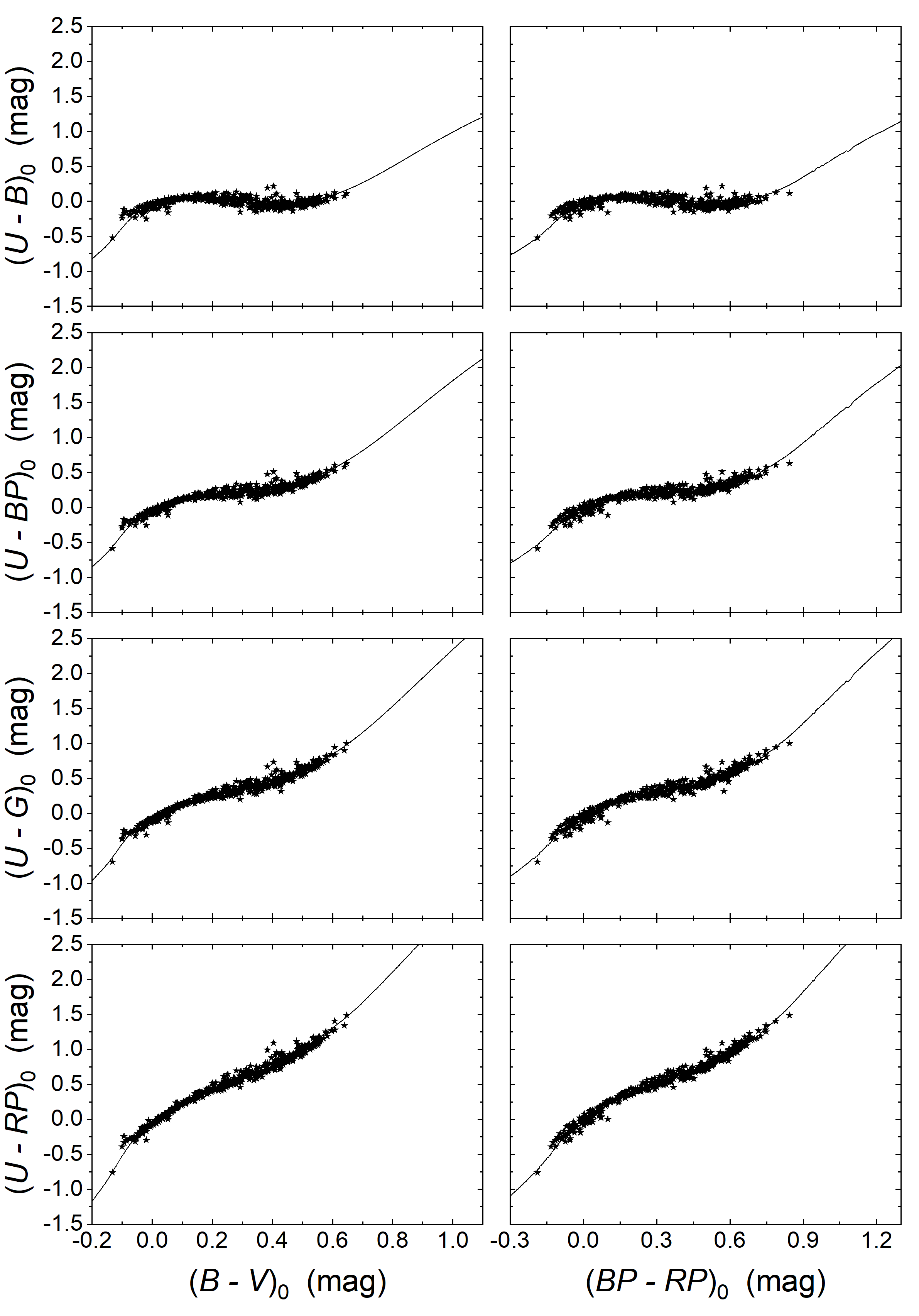}
    \caption{The eight colour-colour diagrams for NGC 1912.
    The x- and y-axis have the same scales. The plots show the different slopes of
    the standard lines.}
    \label{NGC1912}
\end{figure*}

\begin{figure*}
    \centering
    \includegraphics[width = 0.6\textheight]{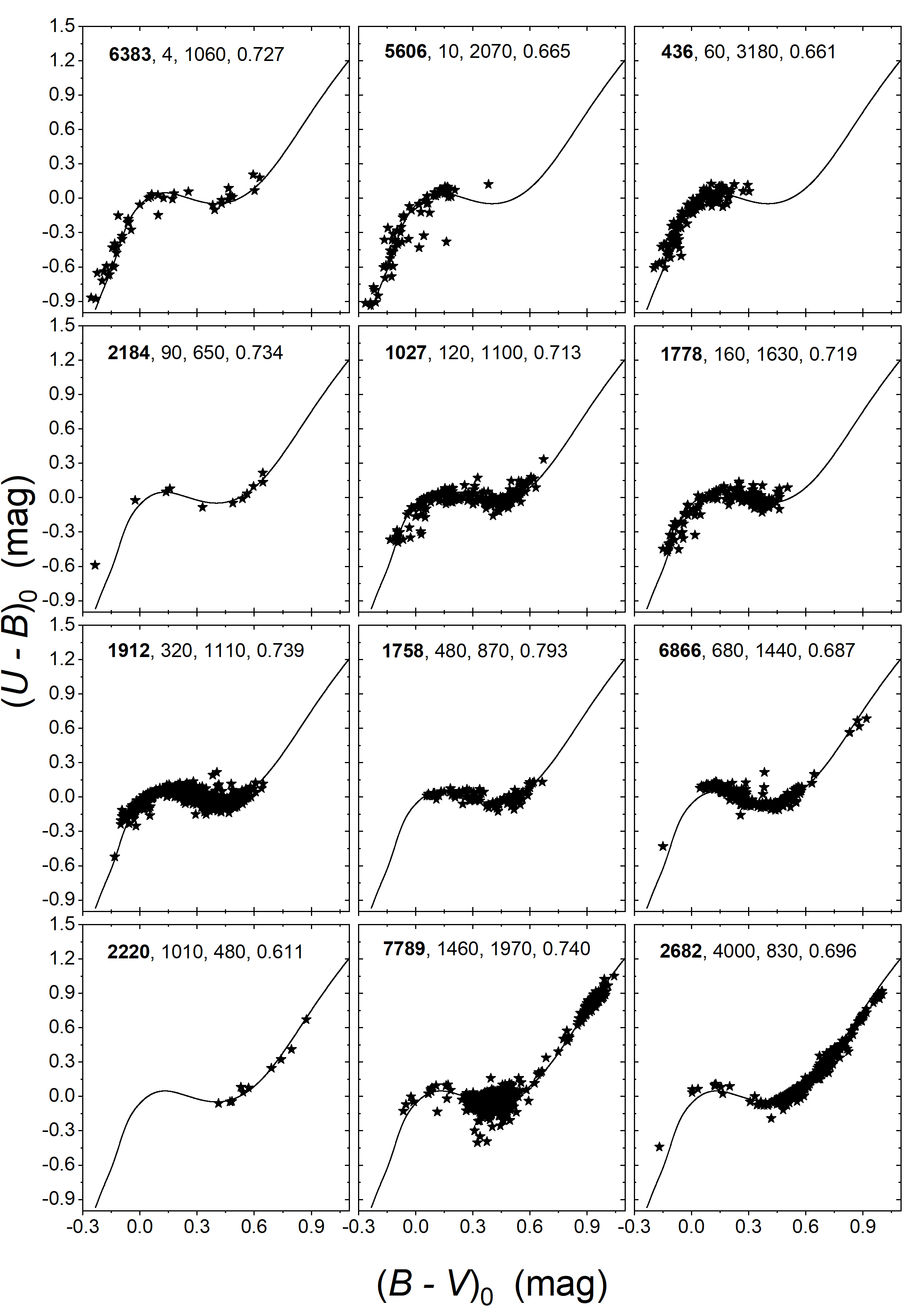}
    \caption{The colour-colour diagrams $(B-V)_\mathrm{0}$ versus $(U-B)_\mathrm{0}$ for
    twelve open clusters sorted by their ages. The values in each panel are the NGC
    number, age (in Myr), distance (in pc) and the ratio $E(U-B)/E(B-V)$. The 
    population of the main-sequence is affected by the age (different
    evolutionary stages of the members) and the distance (magnitude limit of $U$).}
    \label{BV_UB_12_OCLs}
\end{figure*}

\begin{figure*}
    \centering
    \includegraphics[width = 0.6\textheight]{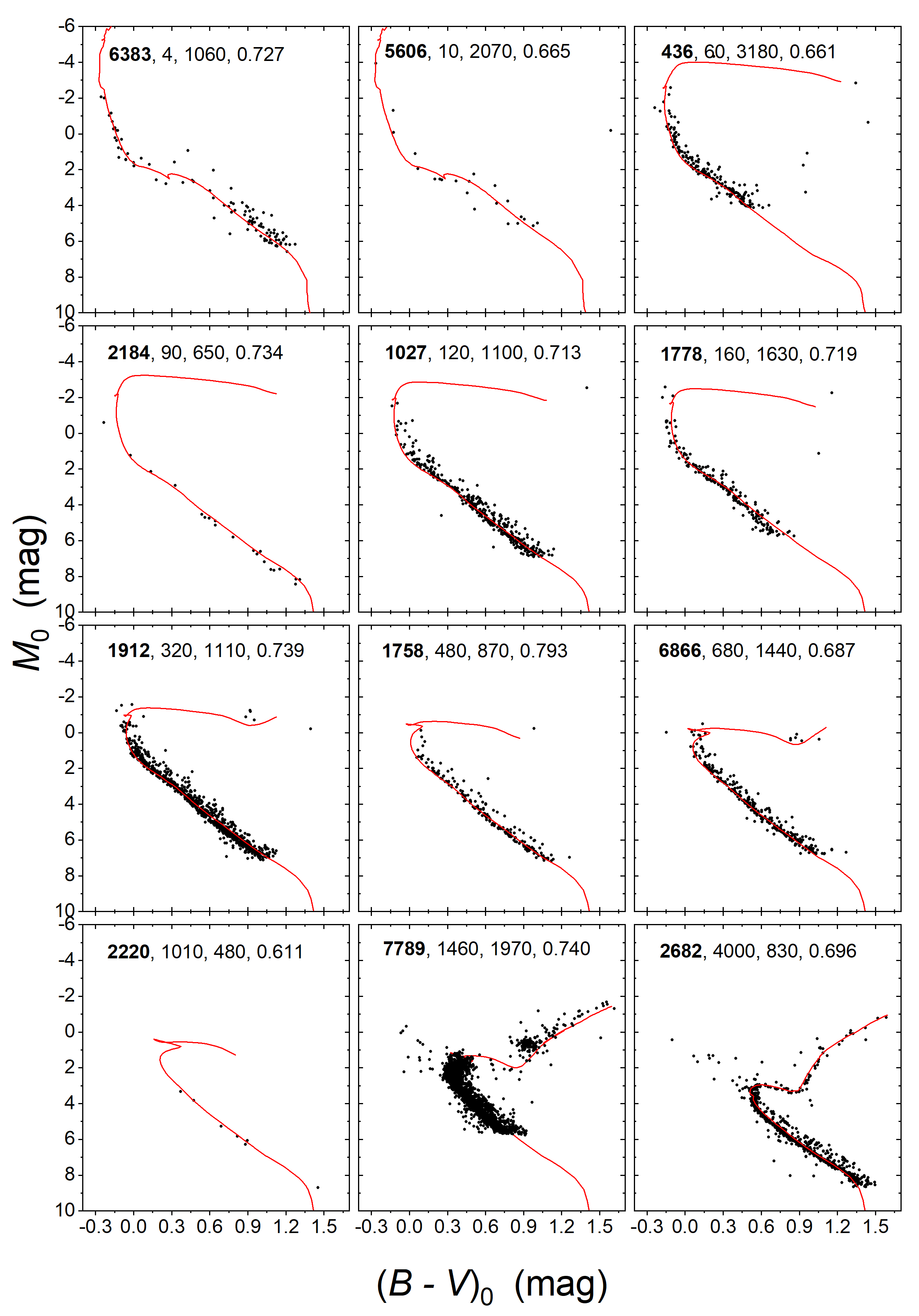}
    \caption{The colour-magnitude diagrams $(B-V)_\mathrm{0}$ versus $M_\mathrm{0}$ for
    twelve open clusters (Fig. \ref{BV_UB_12_OCLs}) sorted by their ages. The values in each panel are the NGC
    number, age (in Myr), distance (in pc) and the ratio $E(U-B)/E(B-V)$. The lines
    are the corresponding isochrones for solar metallicity taken
    from \citet{2025A&A...701A.258N}.}
    \label{BV_MV_12_OCLs}
\end{figure*}

\end{appendix}

\end{document}